\documentclass[fleqn,usenatbib]{mnras}
\usepackage{amsmath}
\usepackage{multirow}
\usepackage{mathptmx}
\usepackage{txfonts}
\usepackage[T1]{fontenc}
\usepackage{enumitem}
\DeclareRobustCommand{\VAN}[3]{#2}
\let\VANthebibliography\thebibliography
\def\thebibliography{\DeclareRobustCommand{\VAN}[3]{##3}\VANthebibliography}

\usepackage{graphicx}	
\usepackage{gensymb}
\usepackage{xcolor}   

\title[Galactic Centre Magnetic Field]{Magnetic Field at the Galactic Centre from Multi-Wavelength Dust Polarization}

\author[M. S. Akshaya and T. Hoang]{
	M. S. Akshaya$^{1}$\thanks{E-mail: akshayams@kasi.re.kr (MSA)}
	and Thiem Hoang$^{1,2}$
	\\
	$^{1}$Korea Astronomy and Space Science Institute, Daejeon 34055, Republic of Korea\\
	$^{2}$Department of Astronomy and Space Science, University of Science and Technology, 217 Gajeong-ro, Yuseong-gu, Daejeon 34113, Republic of Korea\\
}

\date{Accepted XXX. Received YYY; in original form ZZZ}

\pubyear{2024}

\begin{document}
	\label{firstpage}
	\pagerange{\pageref{firstpage}--\pageref{lastpage}}
	\maketitle
	
	\begin{abstract}
		We have mapped the magnetic field ($B$-field) for a region of about 30 pc around the centre of our Galaxy, which encompasses the circumnuclear disk (CND), the minispiral, and the 20 km s$^{-1}$ and 50 km s$^{-1}$ molecular clouds, using thermal dust polarization observations obtained from SOFIA/HAWC+ and JCMT/SCUPOL. We decompose the spectra of $^{12}$CO ($J\!=\!3\!\rightarrow\!2$) transition from this region into individual cloud components and find the polarization observed at different wavelengths might be tracing completely different layers of dust along the line of sight. We use modified Davis-Chandrasekhar-Fermi methods to measure the strength of $B$-field projected in the plane of the sky ($B_{{}_{\mathrm{POS}}}$). The mean $B_{{}_{\mathrm{POS}}}$ of the CND and the minispiral, probed at 53 \micron\ is of the order of $\sim\!2$ mG. $B_{{}_{\mathrm{POS}}}\!\!\!<\!1$ mG close to the Galactic Centre, in the region of the ionized mini-cavity within the CND, and increases outwards. However, the longer wavelength polarization at 216 \micron\ appears to come from a dust layer that is cooler and behind the CND and has a stronger $B$-field of about 7 mG. The $B$-field strength is lowest along the Eastern Arm of the minispiral, which is also the only region with Alfv\'en Mach number, $\mathcal{M}_{\mathrm{A}}>1$ and mass-to-flux ratio, $\lambda\!\gtrsim\!1$. Such an observed weak $B$-field could be a result of the low resolution of the observation, where the tangled $B$-fields due to the strong turbulence in the high density clumps of the CND are lost within the beam size of the observation.
		
	\end{abstract}
	
	
	
	
	\section{Introduction} \label{sec:Intro}
	The centres of galaxies play a vital role in their evolution, right from star formation to quenching. A significant fraction of the galactic star formation might be driven by the inflow of material into the galactic centre, whereas the massive outflows from these starburst activities as well as from the supermassive black hole at the galactic centre (like the Sgr A$^*$ in case of the Milky Way) can trigger galactic quenching and hence its evolution \citep[][and references therenin]{Oort1977,Kormendy2004,Veilleux2020}. Only in our Galaxy do we have the capability to study this complex region in great detail at high resolution using various observational techniques like imaging, spectroscopy, and polarimetry. It is well known that the star formation rate (SFR) of the Milky Way is far below what is expected from gravitational collapse of molecular clouds, and the SFR in the Galactic Centre (GC) is even lower than the average SFR in the Milky Way, yet the origin for the low SFR in the GC remains elusive \citep[][and references therein]{Genzel2010,Barnes2017,Bryant2021,Henshaw2023}. The magnetic field ($B$-field) is one of the popular candidates that can play a role in this suppressed star formation. In order to understand the role of $B$-fields in star formation and evolution of the GC, it is crucial to map its strength and morphology.  
	
	Significant efforts have been made to map the magnetic field of the GC in the past, with most of these studies focused on the circumnuclear disk which is the molecular reservoir closest to Sgr A$^*$ \citep[CND;][]{Becklin1982,Guesten1987,Jackson1993}. Whether the CND hosts active star formation is a matter of debate. There have been indications of ongoing star formation in the CND \citep{Yusef-Zadeh2015} but the evidence is not conclusive \citep{Mills2017}. There are also suggestions that the CND itself is a transient feature like other Galactic Centre clouds \citep{Requena-Torres2012,Mills2017,Dinh2021,Henshaw2023}. Thus, we still do not have a complete picture of the dominant physical conditions of the CND. The line-of-sight (LOS) strength of the $B$-field ($B_{{}_{\mathrm{LOS}}}$) in the CND was estimated from the observations of Zeeman splitting in spectral lines. The average $B_{{}_{\mathrm{LOS}}}$ from the Zeeman measurements at various locations along the CND was about 3 mG \citep{Schwarz1990,Killeen1992,Plante1995,Marshall1995,Yusef-Zadeh1996,Yusef-Zadeh1999}. \citet{Aitken1986} and \citet{Aitken1998} predicted the upper and lower limits of the magnetic field in the region based on the mid-infrared thermal dust polarization observations. They assumed paramagnetic relaxation as the mechanism of grain alignment, and estimated the field to be between 2 -- 10 mG around the GC. However, this assumption no longer holds as recent studies show that paramagnetic relaxation by itself is not strong enough to drive grain alignment \citep{Hoang2016}. Regardless, polarized thermal emission from aligned dust grains is still a popular tool to map the plane-of-sky (POS) magnetic field ($B_{{}_{\mathrm{POS}}}$). This technique is based on the fact that non-spherical dust grains tend to align with their longest axis perpendicular to the $B$-fields \citep{Lazarian2007JQSRT,LazarianRAT2007}, so that the polarization of thermal emission is perpendicular to the $B$-fields \citep{Hildebrand1988}. As a result, by observing the thermal dust polarization and rotating polarization vectors by 90\degree, we can infer the morphology of the POS $B$-field. However, as a caveat, it must be noted that the derived morphologies from this method are biased due to the 180$\degree$ ambiguity intrinsic to the polarization observations, without accurate corrections from the LOS $B$-field measurements. The GC makes an ideal target for mapping the magnetic fields using dust polarization because the dust grains are expected to be efficiently aligned with the $B$-fields in this environment. Due to the crowding of the field in this region, both along the LOS and within the telescope beams, the measured Stokes parameters are an average indicator of the dust grain orientations along these sightlines. Indeed, if the dust grains are not aligned with the magnetic field, then the dust polarization does not trace the $B$-field morphology. However, from our previous study \citep{Akshaya2023}, we have found that, due to the high $B$-field strength observed in the region by Zeeman measurements, grains could achieve perfect alignment through Magnetically-Enhanced Radiative Torque Alignment mechanism \citep[MRAT;][]{Hoang2016,Hoang2022ApJ}. Therefore, dust polarization can be a robust tracer of the $B$-fields in the GC. 
	
	The GC is a complex and dynamic environment known to have densities, pressures, and $B$-fields orders of magnitude greater than those observed in the diffuse interstellar medium (ISM). Molecular spectra of the region also indicate the presence of complex multi-component structures along the LOS \citep{Sutton1990,Henshaw2016,Eden2020,Hu2022}. If these dust components present at various distances along the LOS are emitting radiation at different wavebands (due to the difference in their temperatures), we can use the different polarization morphology observed at multiple wavelengths to get an understanding of the change in the $B$-field strength, morphology, and orientation along the LOS. 
	
	In this paper, we measure the strength of the $B$-field within a region of about 30 pc around the supermassive black hole Sgr A$^*$ at the centre of our Galaxy. We will use polarization observations in three wavebands, the same which were used in our previous study \citep{Akshaya2023} at 53, 216, and 850 \micron. The observations focus on the CND, which is a warm torus of gas and dust orbiting around Sgr A$^*$, and the minispiral which encompasses a set of ionized gas filaments present within and interacting with the CND \citep{Lo1983,Christopher2005}. Each observation probes the region on different scales allowing us to map the small and large scale magnetic field in the region. The CND is covered in all three observations while part of the 20 km s$^{-1}$ and 50 km s$^{-1}$ clouds \citep{Kauffmann2017} are seen in the 216 \micron\ observation. The 850 \micron\ data covers the largest area around Sgr A$^*$ which includes the CND, minispiral, 20 km s$^{-1}$, and 50 km s$^{-1}$ clouds. A recent study of the CND using Stratospheric Observatory for Infrared Astronomy \citep[SOFIA;][]{Temi2018} observation by \citet{Guerra2023} estimate the $B$-field strength of the minispiral to have median values between 5 -- 8 mG on a spatial scale of $\lesssim1$ pc. 
	
	The most commonly used technique for the measurement of the POS magnetic field strength from thermal dust polarization is the Davis-Chandrasekhar-Fermi (DCF) method \citep{Davis1951,Chandrasekhar1953}. It is based on the propagation of Alfv\'en waves through the medium, when there is an energy balance between the gas turbulence and magnetic turbulence. Assuming the magnetic field lines to be frozen with the matter in the general ISM conditions, the method attributes any small scale irregularities observed in the polarization vectors to turbulence. The net magnetic field in the region can be assumed to be made up of a regular component ($B_0$) and a turbulent component ($\delta B$). A strong $B_0$ resists being perturbed by turbulence such that $\delta B\ll B_0$, thus allowing us to characterise the POS component of $B_0$ ($B_{{}_{\mathrm{POS}}}$) by measuring the net irregularity of the field lines using the relation,
	\begin{equation} \label{eq:1}
		B_{{}_{\mathrm{POS}}} = \sqrt{4\pi\rho}\frac{\sigma_v}{\sigma_{\phi}},
	\end{equation}
	where $\rho$ is the gas mass density, $\sigma_v$ is the turbulence-induced velocity dispersion measured from non-thermal line-width measurements, and $\sigma_{\phi}$ is the distortion in the $B$-field measured by the polarization angle dispersion. Various studies have tested the validity of the method in estimating the POS $B$-field using 3D numerical magnetohydrodynamic (MHD) simulations \citep[][and references therein]{Ostriker2001,Heitsch2001,Falceta2008,Liu2021,Chen2022,Myers2024}. They found the DCF method to predict the POS $B$-field with considerable accuracy (within a factor of 2), especially in strong field environments on scales $\gtrsim0.1$ pc. However, in spite of its wide usage, the method overestimates the magnetic field due to its restrictive initial assumptions about the conditions of the underlying medium \citep{Hildebrand2009,Houde2009,Cho2016,Chen2022} and underestimates the $B$-field if it is derived based on compressible non-Alfv\'en modes \citep{Skalidis2021,Myers2024}. Thus, the DCF method is a useful technique to get an average estimate of the magnetic field in simple environments without self-gravity or shear motion.
	
	The DCF method has been applied successfully by many polarization studies \citep{Pillai2015, PlanckXXXV2016, Pattle2017, Guerra2021, Ngoc2021, Hwang2021, HoangM172022}. Some of the physical conditions that contribute to the over-estimate of the mean $B_{{}_{\mathrm{POS}}}$ while using the DCF method include; anisotropic turbulence, failure of equipartition between the magnetic and kinetic energy, and self gravity. It is difficult to disentangle the importance of each of them without a detailed understanding of the kinematics of the region. The overestimation could also be a result of integration effects from within the beam size of individual observations, as well as along the LOS. These uncertainties arise from the polarization observation and impact the parameter $\sigma_{\phi}$ in Equation \ref{eq:1} \citep{Hildebrand2009, Houde2009, Skalidis2021, Li2022, Guerra2023}. However, \citet{Chen2022} found that the hydrodynamic properties of gas in the region contribute equally if not more to the uncertainty of the measured field (reflected in the parameters $\rho$ and $\sigma_v$). 
	
	\begin{figure*}
		\centering
		\includegraphics[scale=0.4]{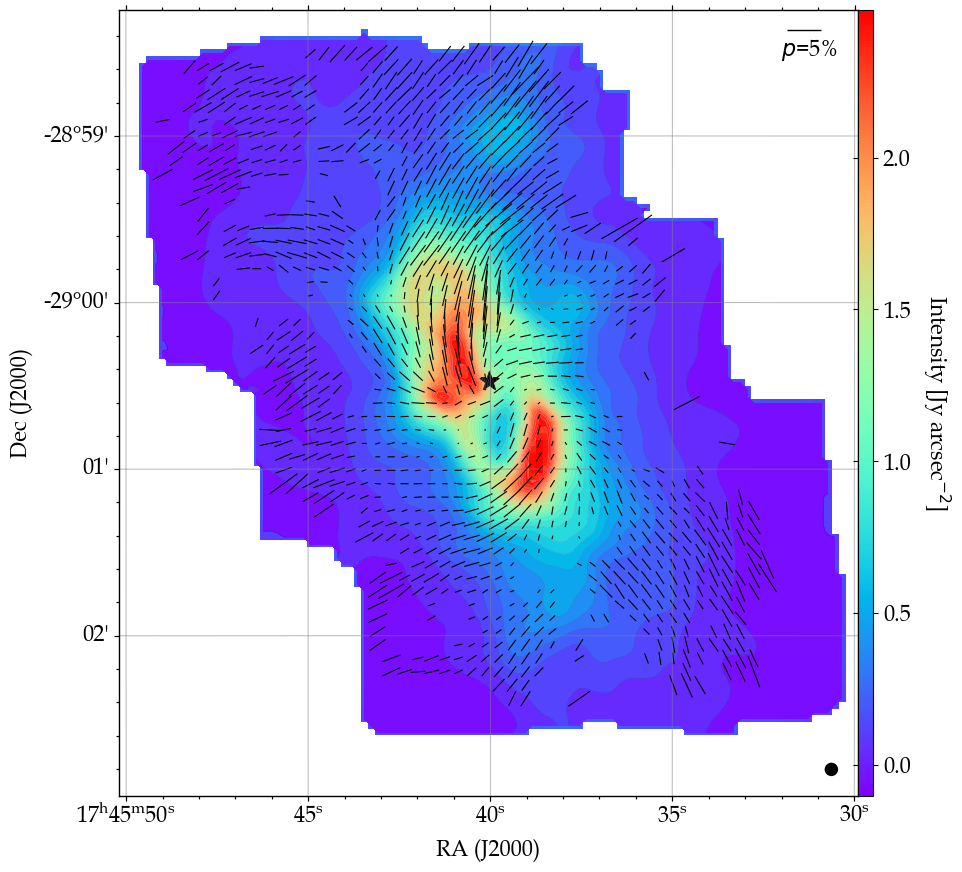}
		\hspace{3cm}
		\includegraphics[scale=0.4]{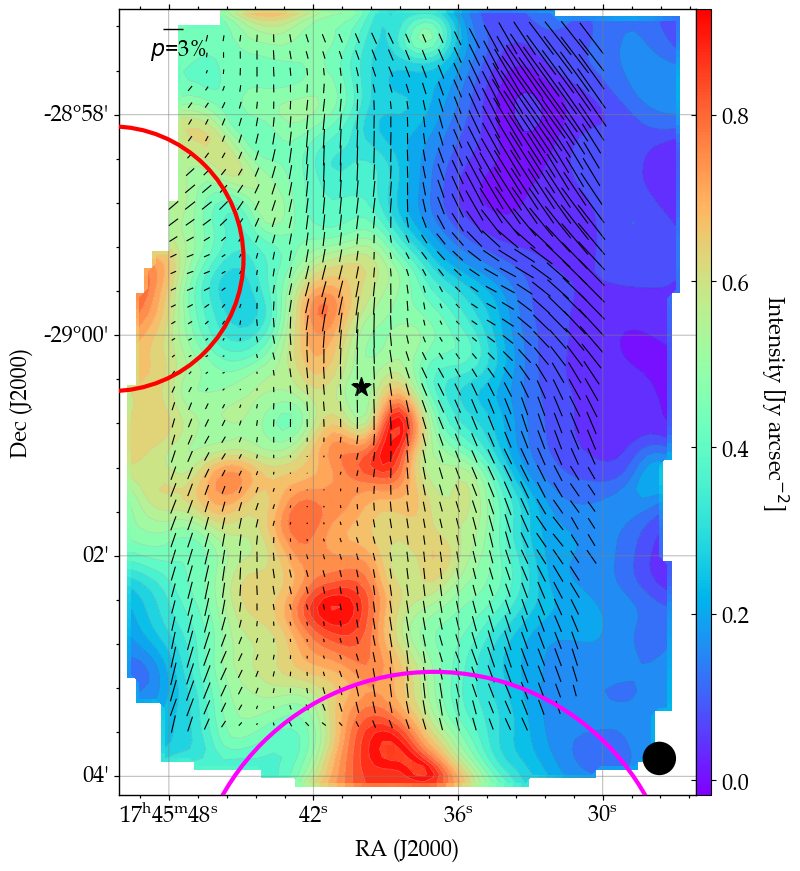}
		\includegraphics[scale=0.4]{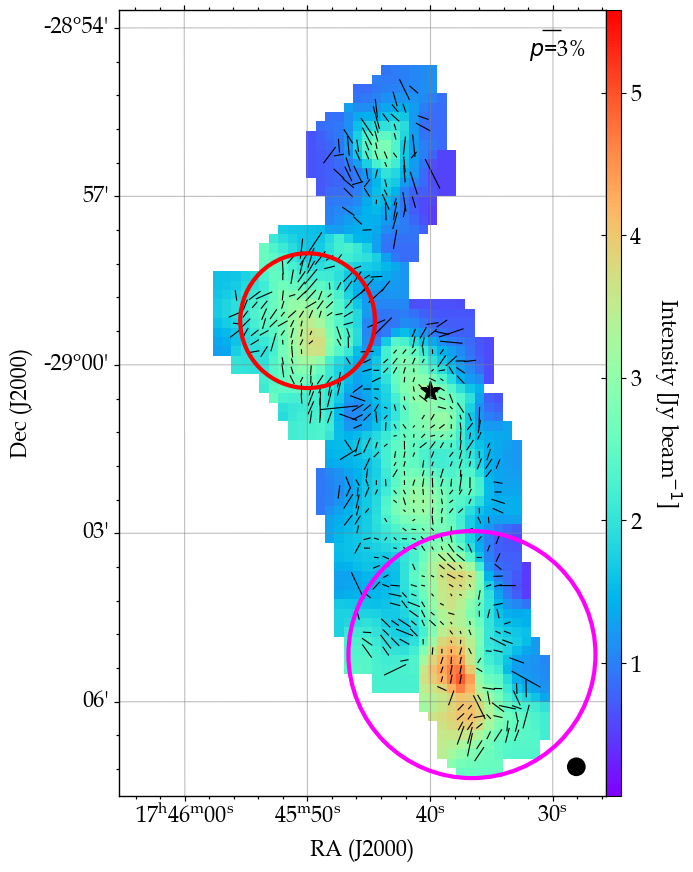}
		\caption{The maps of polarization at 53 \micron\ (top) and 216 \micron\ (bottom left) from SOFIA/HAWC+, and at 850 \micron\ (bottom right) from  JCMT/SCUPOL. The colorbars represent the intensity in respective wavebands and the beam size of each instrument is shown on the bottom right of each figure along with the representative scale of polarization percentage at the top. The star symbol in all the images marks the position of Sgr A$^*$. The red and magenta circles in the 216 \micron\ and 850 \micron\ maps indicate the positions of the 50 km s$^{-1}$ and 20 km s$^{-1}$ clouds, respectively.}
		\label{fig:pol_maps}
	\end{figure*}
	
	Several attempts have been made to improve the original DCF technique. The DCF method is a good approximation when the observed angle dispersion is small i.e. when $\delta B\ll B_0$ \citep{Ostriker2001}. \citet{Falceta2008} extended the DCF method to cases where the turbulent component of the magnetic field is comparable to the mean field, thus resulting in large angle dispersion. \citet{Hildebrand2009} and \citet{Houde2009} further improved the DCF estimate by using the turbulent-to-ordered magnetic field ratio to estimate the angel dispersion, based on the second order structure function of magnetic field position angles introduced by \citet{Falceta2008}. This method incorporated the large scale structure of the magnetic field along with the instrumental effects in the observations. The recent modifications include those by \citet{Cho2016} and \citet{Lazarian2022} where they address the anisotropic nature of MHD turbulence, by incorporating structure function in combination with velocity centroids to estimate the velocity fluctuations in the POS. We will apply the DCF modifications from \citet{Houde2009} and \citet{Lazarian2022} in our study, to understand how multi-wavelength polarization can be used to probe the strength of the 3D $B$-field (i.e. $|B|=(B^2_x+B^2_y+B^2_z)^{1/2}$). No correction factors are used in our estimates and they can be considered as an upper limit of the $B$-field in the region covered by each observation.
	
	Our initial study of the thermal dust polarization around the GC is discussed in \citet{Akshaya2023}, where we focus on the grain alignment physics. We have used the same observations in this work to understand how the strength of the magnetic field varies across different scales and wavelengths along the LOS, with longer wavelengths probing deeper into the LOS compared to shorter wavelengths. The region of the Galactic disk in general, and GC in particular are known to have multiple structures at different distances along the LOS. We will use recent techniques to isolate the individual components and understand their effects on the derived magnetic field.
	
	The rest of our paper is structured as follows; Section \ref{sec:Observations} describes the polarization observations and their data quality assessment. The estimation of the gas velocity dispersion is discussed in Section \ref{sec:Velocity_Dispersion}. The magnetic field measured from the DCF modifications by \citet{Houde2009} and \citet{Lazarian2022} is described in Section \ref{sec:B_field}. A detailed discussion of the estimated $B$-field and its implications on the kinematics of the regions are presented in Section \ref{sec:Discussion}, followed by a brief summary of our results in Section \ref{sec:Summary}.
	
	\begin{figure}
		\centering
		\includegraphics[scale=0.38]{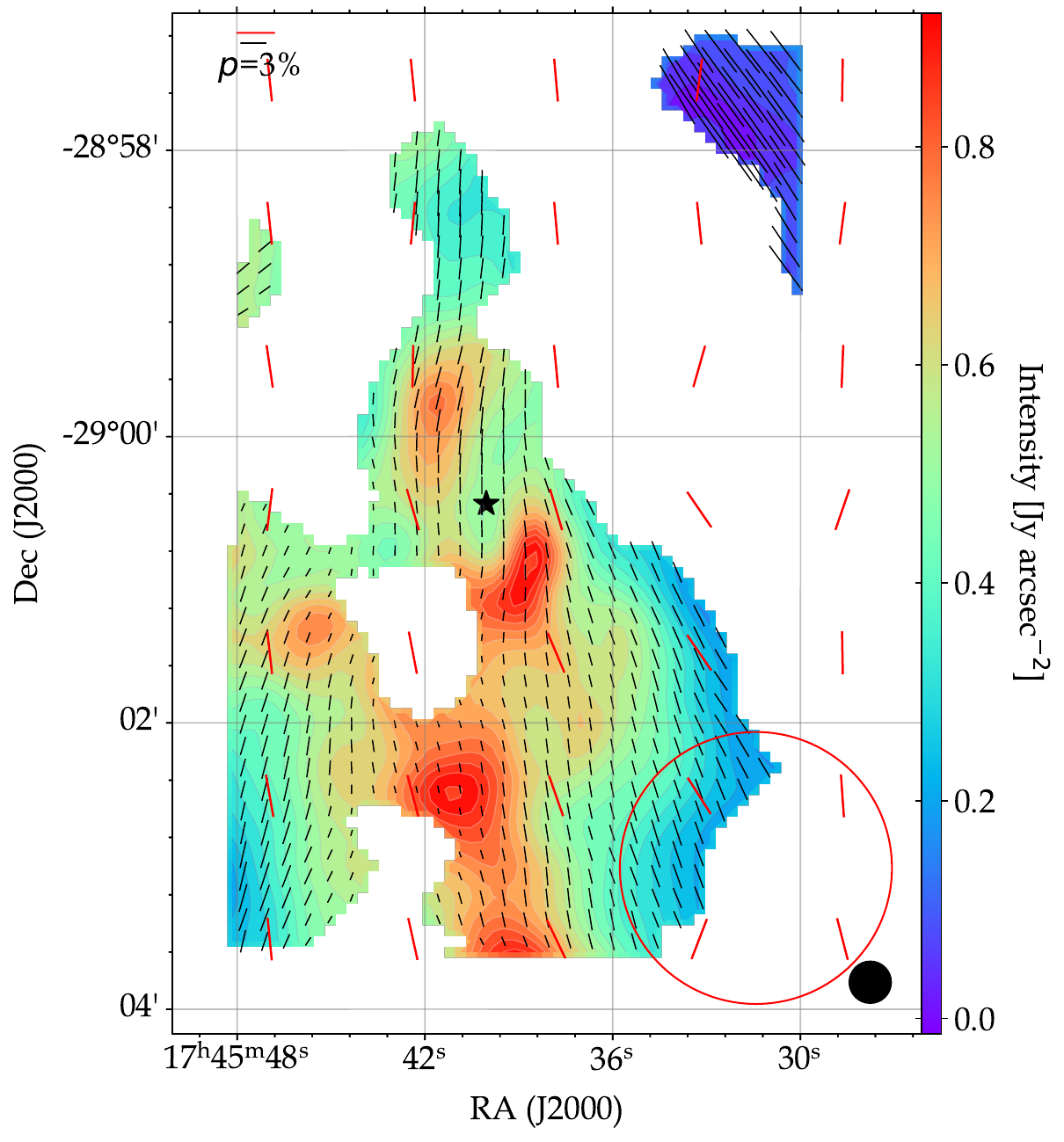}
		\includegraphics[scale=0.38]{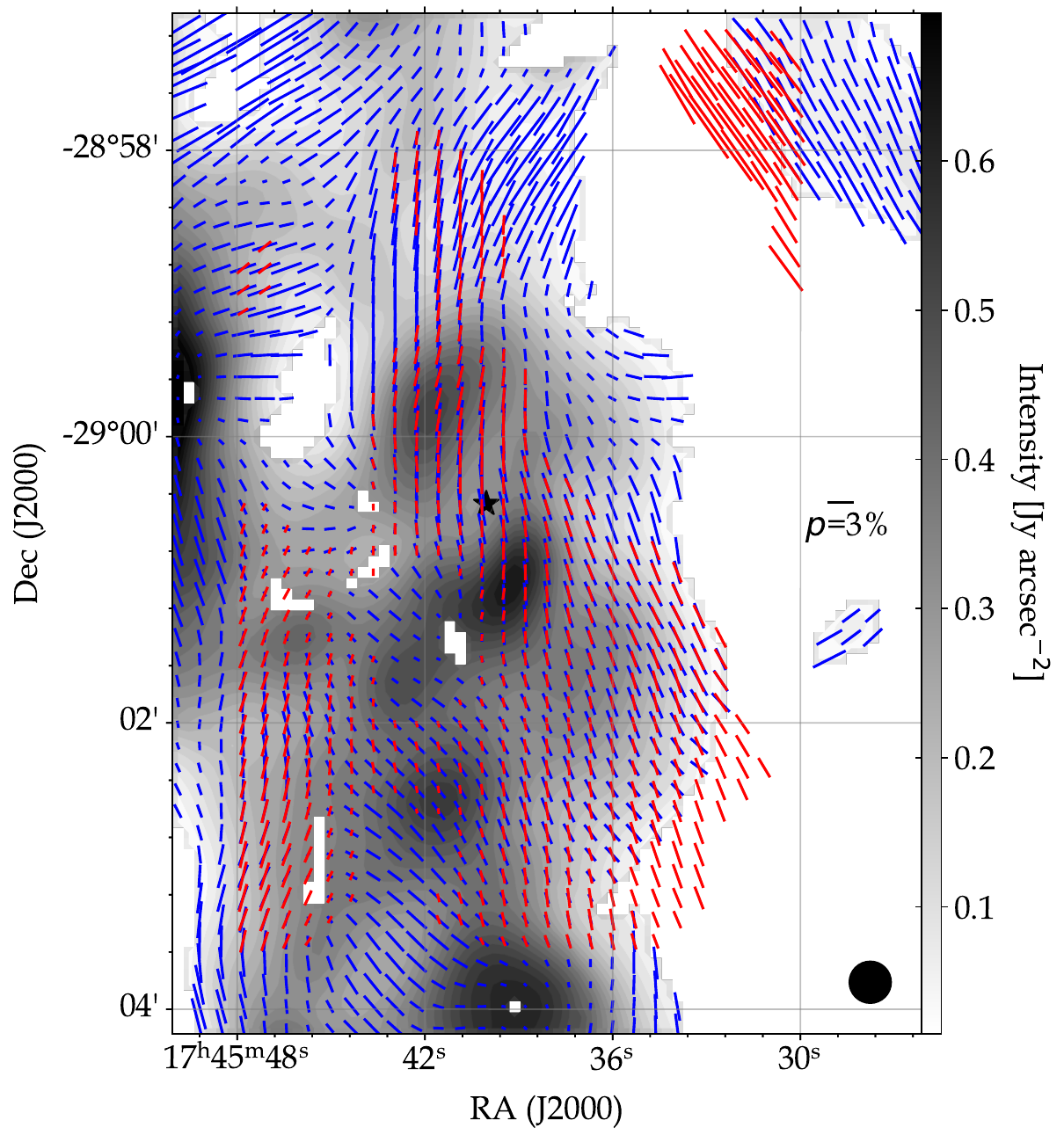}
		\caption{The top map shows the HAWC+ polarization observation of the CND at 216 \micron\ with the criteria $\Delta\phi_{\mathrm{ref}} < 10\degree$, compared with the 240 \micron\ polarization from PILOT (red vectors). The background intensity map is the Stokes I measurements from HAWC+. The PILOT polarization vectors are scaled to $p=3\%$. The map at the bottom compares the HAWC+ observations of the same region taken using the CNM (red) and OTFMAP (blue) mapping strategies respectively. The background intensity in this case is from the Stokes I measurements of the OTFMAP. The scale of polarization percentage is shown on each map along with the beam sizes of the HAWC+ (in both images) and PILOT (top image) instruments.}
		\label{fig:s216-pilot}
	\end{figure}
	
	\section{Dust Polarization Observations} \label{sec:Observations}
	We have used the thermal dust polarization observations around the GC at 53 and 216 \micron\ from the High-resolution Airborne Wide-band Camera Plus \citep[HAWC+;][]{Harper2018}, which was the far-infrared imager and polarimeter for NASA’s Stratospheric Observatory for Infrared Astronomy \citep[SOFIA;][]{Temi2018} and the reprocessed data at 850 \micron\ from SCUPOL, which was the polarimeter for the Submillimeter Common User Bolometer Array (SCUBA) instrument on the James Clerk Maxwell Telescope (JCMT) presented by \citet{Matthews2009}. More details about the data and the polarization cut-off criteria used are presented in \citet{Akshaya2023}, where the grain alignment of the region was studied with the same observations. The SOFIA/HAWC+ observations were reduced using the latest version of the HAWC+ data reduction pipeline i.e. DRP v3.2.0 compared to the DRP v1.3.0 used in our previous analysis. The instrumental polarization (IP) in the reduced Stokes parameters (i.e. $q=Q/I$ and $u=U/I$) is subtracted during the data reduction procedure. These values ($q_{_{IP}},u_{_{IP}}$) are calculated from measurements of the sky for each filter band and correspond to $(-0.0157,-0.0038)$ at 53 \micron\ and $(-0.0104,-0.0142)$ at 216 \micron. The quality of these datasets are nominal with all the problematic files removed during the reduction procedure. 
	
	We have also used the data from \textit{Herschel} in five wavebands at 70, 160, 250, 350, and 500 \micron\ to derive the gas column density ($N_{{}_\mathrm{H}}$) and the dust temperature ($T_{\mathrm{d}}$) as described in \citet{Akshaya2023}. The polarization maps of each observation is shown in Fig. \ref{fig:pol_maps}. The magnetic field strength will be estimated for each of the polarization observations, which we believe to be probing different components of dust along the LOS. Throughout this work we assume the distance to the GC to be 8.34 kpc \citep{Reid2014}.
	
	\subsection{Data Quality Assessment}
	The observed polarization from the CND and its surrounding can be subject to reference beam contamination due to the emission from extended features in this region. This is more of a problem in the 216 \micron\ observation due to the temperature of the surrounding material. At 53 \micron\ the dust being probed is much hotter than its surrounding and hence the contribution from the reference beams is not as significant, allowing us to use the Level 4 data products as it is at this wavelength. We have estimated the level of reference beam contamination in the SOFIA/HAWC+ 216 \micron\ observation using the method described by \citet{Novak1997} and \citet{Chuss2019}. The chop angle and amplitude of the observation are 60$\degree$ and 240$''$, respectively with background images taken from two chop positions symmetrically on either side of the source. \textit{Herschel} 70, 160, 250, and 350 \micron\ observations of the region were used to model the expected intensity of the reference beams in the HAWC+ filter band-pass using the relation;
	\begin{equation}
		I = A\nu^2B_{\nu}(T_\mathrm{d}),
	\end{equation}
	where $A$ is the amplitude, $\nu$ is the frequency, and $B_v(T_{\mathrm{d}})$ is the Planck function corresponding to the dust temperature $T_{\mathrm{d}}$. The mean intensity of the reference beams was used in combination with the calibrated Stokes I intensity from HAWC+ to determine the contrast of the on-target beam with respect to the reference beams. Due to the far-infrared brightness of the GC region, the ratio of these intensities (represented by $w\equiv I_r/I_m$, where $I_r$ is the average intensity from the reference beams and $I_m$ is the measured Stokes I intensity of the polarization observation) was found to be $I_m<6I_r$ throughout the region. The established method is to only consider polarization vectors where the contrast is greater than 10 \citep{Santos2019}. However, this seems unlikely in the Galactic disk at these temperatures due to the presence of extended emissions along most of the lines of sight. The other approach is to quantify the level of contamination in the measured polarization fraction ($p$) and polarization angle ($\phi$). Since our goal is to measure the strength of the $B$-field, we are only interested in the level of contamination in the polarization angle, which is a key parameter to estimate the $B_{{}_{\mathrm{POS}}}$. 
	
	\citet{Novak1997} derived the maximum error in the measured polarization angle with only the intensity estimates from the reference beam to be,
	\begin{equation}
		\Delta\phi_{\mathrm{ref}} = \frac{1}{2} \mathrm{tan}^{-1}\left[\frac{p_r w}{(p_m^2 - p_r^2 w^2)^{1/2}} \right],
	\end{equation}
	where $p_m$ is the measured polarization fraction (without debias) and $p_r$ is the assumed polarization of the reference beam. From Fig. \ref{fig:pol_maps} it can be seen that except the top right region of the 216 \micron\ polarization map, most of the map has a polarization of roughly $p\sim1\%$. This is also the level of the observed polarization in the regions of the reference beams from $Planck$ observations \citep{PlanckXII2020,PlanckLVII2020}. Thus the reference beam polarization was set to $p_r=1\%$ to estimate $\Delta\phi_{\mathrm{ref}}$. Following the cut-off criteria defined in \citet{Chuss2019}, we use only the polarization vectors with $\Delta\phi_{\mathrm{ref}} < 10\degree$ and the resulting polarization map is shown in Fig. \ref{fig:s216-pilot} (top). Earlier estimates of the contamination from reference beam use $p_r=10\%$ \citep{Chuss2019,Lee2021}. However, this value seems too high for our observation and results in the rejection of all the polarization vectors.
	
	Due to the low contrast between the source and reference beams in this region, we also tested the agreement between the polarization angles observed by HAWC+ and those from PILOT balloon experiment presented by \citet{Mangilli2019}, which measured the polarization from an extended region of the GC at 240 \micron\ (with a beam size of $1.9'$). The cut-off based on $\Delta\phi_{\mathrm{ref}}$ removed the vectors from low-intensity regions, which predominantly is the low-intensity part in the right of Fig. \ref{fig:pol_maps}. From our previous analytical study presented in \citet{Akshaya2023}, this was the region where the environmental conditions favour efficient grain alignment, and hence where we can expect the highest degree of polarization. The other removed vectors are from the regions where the observed polarization was very low ($p<0.1\%$). These seem to be the most affected by the reference beam contamination. The polarization vectors which did not satisfy the cut-off set by $\Delta\phi_{\mathrm{ref}}$ also do not match well with the polarization vectors from PILOT observations, especially in the regions of low polarization fraction. We consider the remaining vectors good enough for further analysis and use them as is for the subsequent discussions presented in this paper. 
	
	We have also compared the HAWC+ 216 \micron\ observation after the quality cuts mentioned above with the observations from the same instrument but taken with a different observing strategy as part of the Far-Infrared Polarimetric Large Area CMZ Exploration survey \citep[FIREPLACE;][]{Butterfield2024a,Butterfield2024,Pare2024}. FIREPLACE survey used the on-the-fly mapping mode (OTFMAP) to measure the polarization of the Central Molecular Zone (CMZ) to overcome some of the challenges in observing extended emission features posed by the standard observing strategy of the HAWC+ instrument which was Chop-Nod-Match \citep[CNM;][]{Harper2018}. The main difference between the two strategies lies in the background subtraction procedure. While the CNM observations are limited by background reference images taken close to and symmetrically on either side of the source, the OTFMAP can choose a background region completely devoid of any structures associated with the source, as the scans can begin at a reasonable distance from the target location. This can be very useful when dealing with extended emission features like in the GC region as the reference images necessary for the background subtraction tend to have some remnant features of the extended object, resulting in significant reference beam contamination. The polarization vectors from both observing strategies are shown in the bottom map of Fig. \ref{fig:s216-pilot}, where the OTFMAP map follows the standard cuts to the data with polarization signal-to-noise $p/\sigma_p>3$, polarization degree $p<50\%$, and intensity signal-to-noise (Stokes I) $I/\sigma_I>200$ and the CNM observation follows the same data cut along with $\Delta\phi_{\mathrm{ref}}<10\degree$ in the absence of reference-beam contamination. The main problematic region is the low intense part in the right of the figure where both the OTFMAP and the CNM observations do not retain any polarization vectors. The other regions where we have measurements from both strategies seem to agree well except in the few parts where the CNM observation show a low polarization fraction. Considering the overall agreement in the cut-off regions, we will henceforth use the CNM observation with the $\Delta\phi_{\mathrm{ref}} < 10\degree$ criteria for further analysis.
	
	The 850 \micron\ observation used in this work is a mosaic created from 69 original observations from the SCUPOL archive \citep{Matthews2009}. The archival data were corrected for instrumental polarization of $\sim1\%$ using the procedure outlined in \cite{Greaves2003} and smoothed to an effective beamwidth of $20''$. An estimation of possible contamination in this data is beyond the scope of the current paper. We will use this observation as is for our analysis. 
	
	\begin{figure}
		\centering
		\includegraphics[scale=0.4]{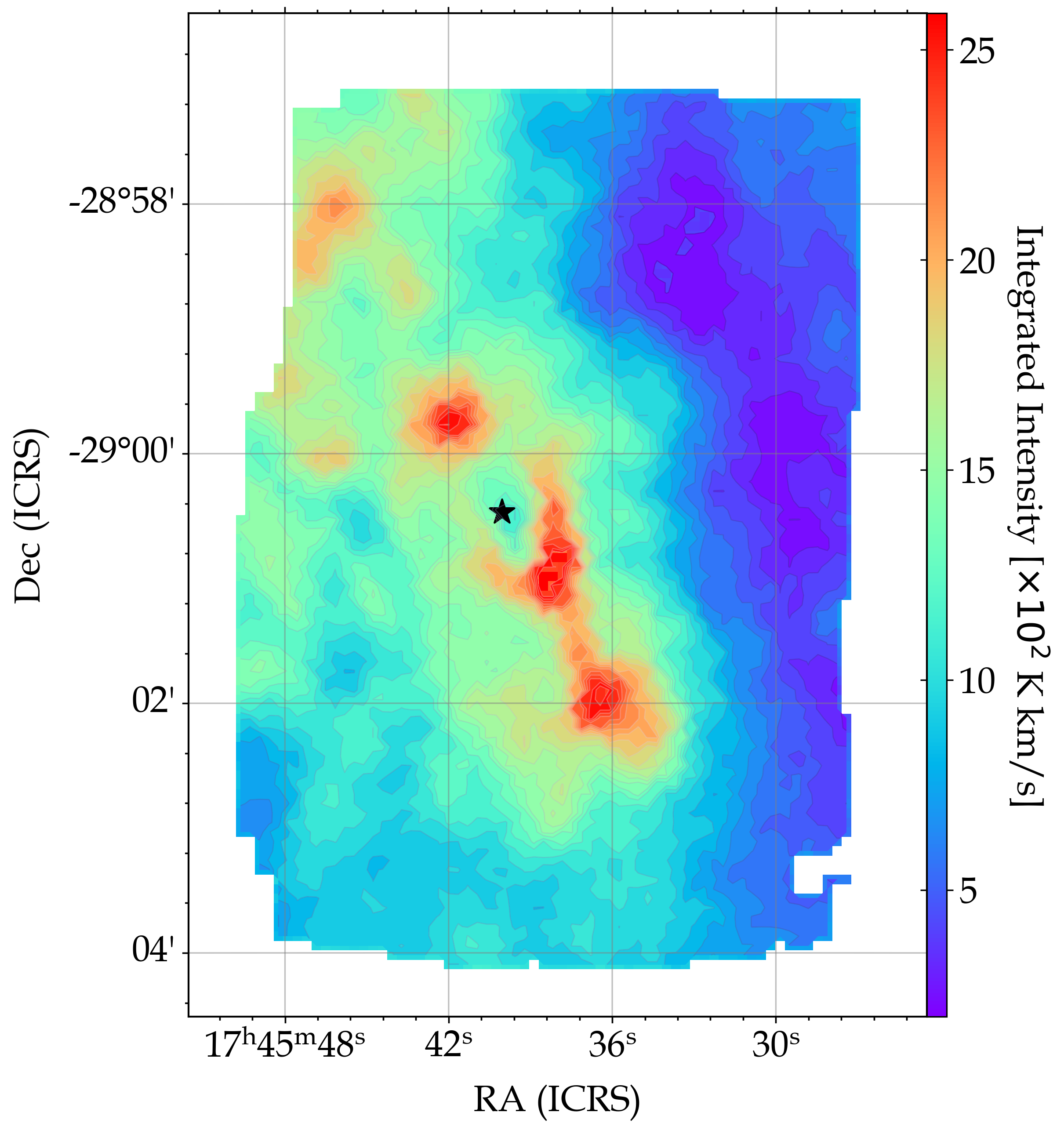}
		\caption{Moment zero map of the CO ($J = 3\rightarrow2$) transition used for our analysis taken from CHIMPS2. The data was smoothed to match the resolution of the SOFIA/HAWC+ 216 \micron\ observation.}
		\label{fig:CO_m0}
	\end{figure}
	
	\begin{figure}
		\centering
		\includegraphics[scale=0.5]{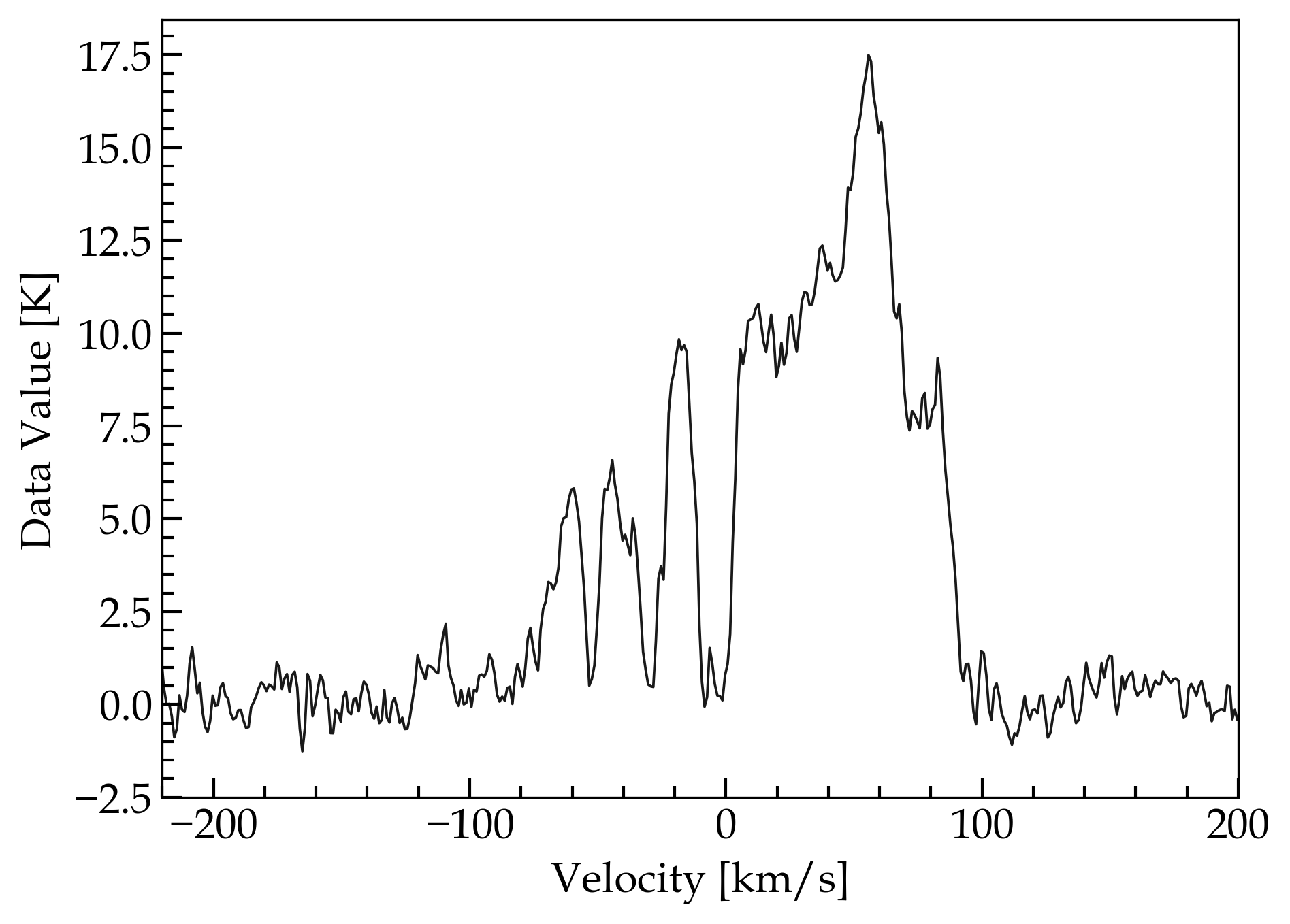}
		\includegraphics[scale=0.5]{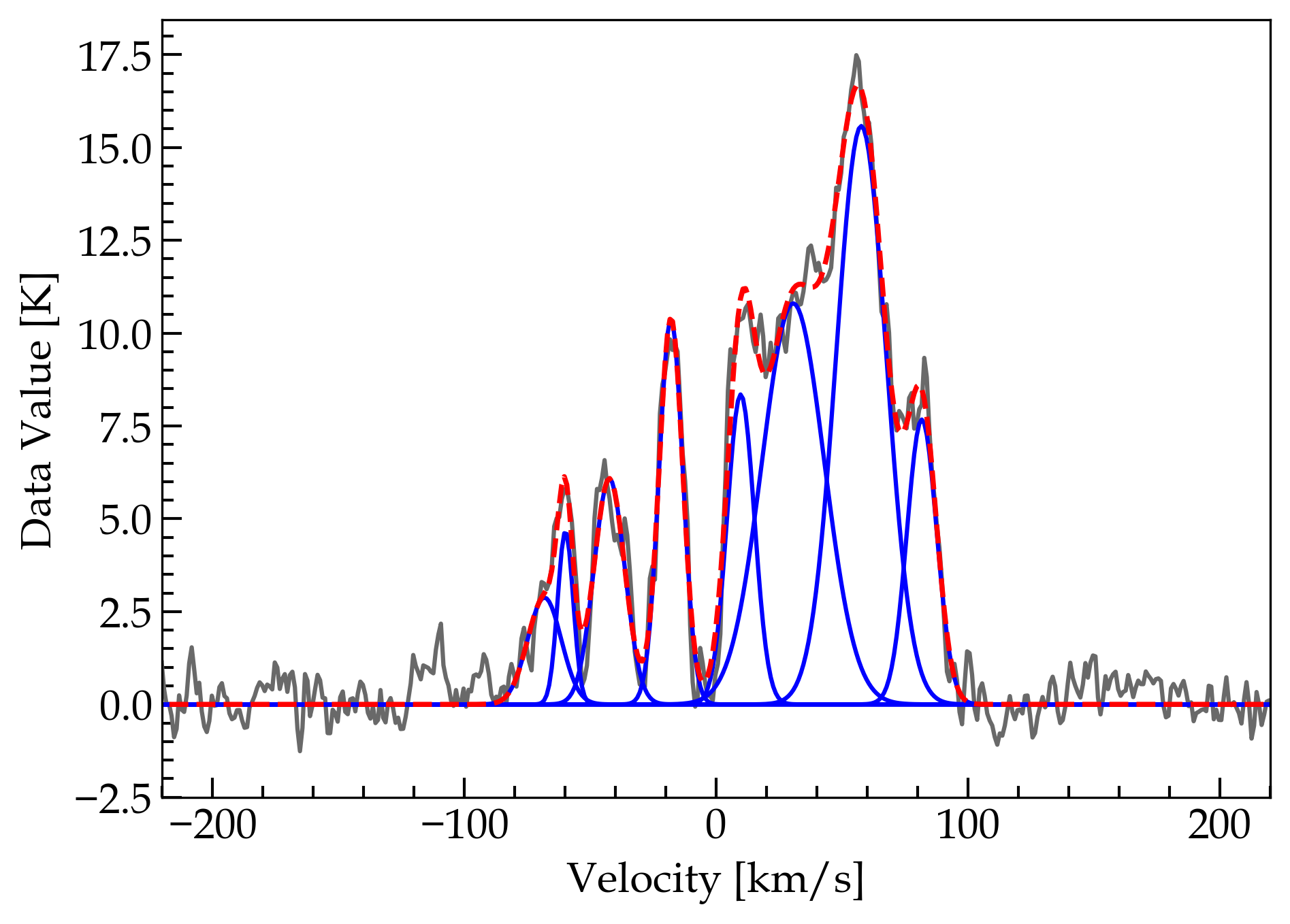}
		\caption{Sample spectrum of the multiple components observed along the GC is shown in the top figure along with the respective Gaussian fits to each identified component in the bottom figure. The blue lines are the decomposed Gaussian components from \textsc{scousepy}. The dotted red line indicate the best fit spectrum from the combination of the decomposed components.}
		\label{fig:sample_spectrum}
	\end{figure}
	
	\section{Gas Velocity Dispersion} \label{sec:Velocity_Dispersion}
	Carbon monoxide (CO) is the second most abundant molecule in the ISM after H$_2$ and is ubiquitous in the Galactic plane, making it an ideal tracer of the velocity structures and the morphology of gas along this LOS. Its rotational transitions can be observed in the submillimeter wavelengths, with different transitions probing different densities. $^{12}{\rm C} ^{16}{\rm O}$ (commonly denoted as CO) is the most common isotopologue of CO and can be used to trace gas densities of the order of $n_{{}_{\mathrm{H}}}\sim10^3-10^4$ cm$^{-3}$ depending on the transition. Other isotopologues like $^{13}{\rm CO}$ and ${\rm C} ^{18}{\rm O}$ have a lower optical depth and can trace denser regions due to their lower abundance than CO. The CO ($J = 3\rightarrow2$) transition is optically thin compared to its other rotational transitions ($J = 1\rightarrow0$ and $J = 2\rightarrow1$) and also has a higher resolution due to its high frequency ($\nu=345.8$GHz). Thus CO ($J = 3\rightarrow2$) can trace warmer and denser environments compared to its lower energy counterparts. Due to the nature of the CO ($J = 3\rightarrow2$) transition, the observations of this line along the Galactic plane often shows multiple complex emission features corresponding to various structures at different distances along the LOS. This can be a drawback in cases where the focus is on a single filament or star forming region and it might be ideal to use higher density tracers like $^{13}{\rm CO}$ or ${\rm C} ^{18}{\rm O}$ for these cases. For our study, we chose to use CO ($J=3\rightarrow2$) transition because we are interested in seeing all the components along this LOS that might be contributing to the observed polarization. 
	
	\begin{figure*}
		\centering
		\includegraphics[clip,trim=0 6cm 0 6cm, scale=0.4]{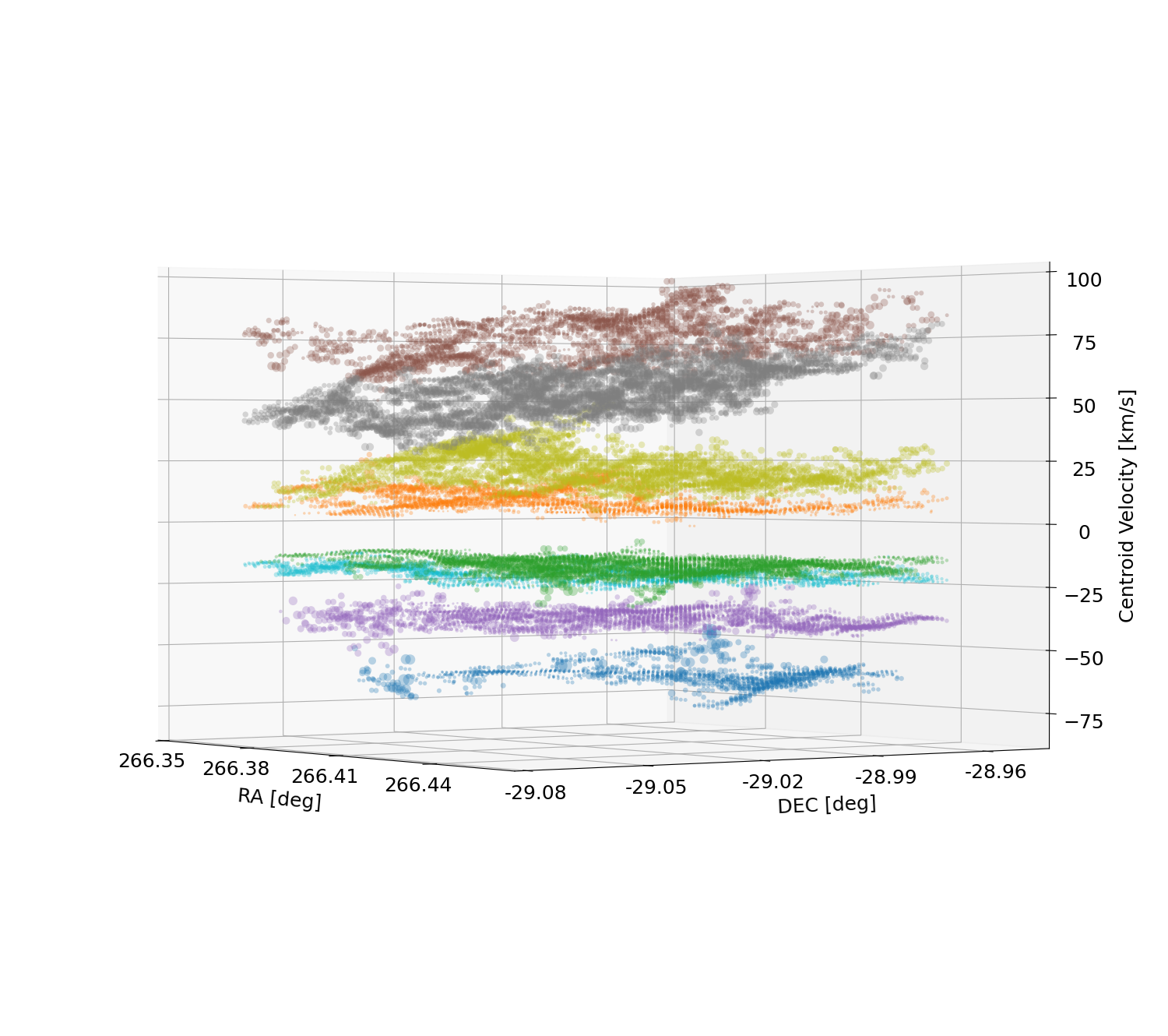}
		\caption{Velocity components observed for the region covered by SOFIA/HAWC+ 216 \micron\ observation, decomposed using \textsc{scousepy} and grouped using \textsc{acorns}. The centroid velocity of the velocity components is along the z-axis and the data points are scaled according to their corresponding velocity dispersion. Each colour represents individual trees resulting from the clustering analysis. Only the trees with a number of leaves greater than four are displayed here.}	
		\label{fig:acorns}
	\end{figure*}
	
	We have used CO Heterodyne Inner Milky Way Plane Survey 2 \citep[CHIMPS2;][]{Eden2020} data for the analysis and estimation of the velocity dispersion along the LOS to the GC. The aim of the CHIMPS2 survey is to map the Inner Galaxy, the CMZ, and a section of the Outer Galaxy in the $^{12}$CO, $^{13}$CO, and C$^{18}$O ($J = 3\rightarrow2$) emissions using the Heterodyne Array Receiver Program (HARP) on the JCMT. We have used the first look data of the CHIMPS2 survey towards the CMZ in CO transition $J=3\rightarrow2$ presented by \citet{Eden2020}, with a spatial resolution of $15''$ and a velocity resolution of 1 km s$^{-1}$. We smoothed the spectral cube to match the spatial resolution of SOFIA/HAWC+ 216 \micron\ and the JCMT/SCUPOL 850 \micron\ observations. The 53 \micron\ observation from HAWC+ is at a much higher resolution than the CO data, hence we have used the derived velocity dispersion values at the resolution of the 216 \micron\ observation as an approximation for its $B$-field calculation. The moment zero map of the data is shown in Fig. \ref{fig:CO_m0}. From the figure it can be seen that the morphology of the prominent features observed in the polarization maps are also observed in the integrated intensity map of the CO spectra. This in general is a good indicator that the chosen spectra can trace the distribution of matter corresponding to the observed polarization along the LOS and is the basic test before choosing any molecular species to measure the turbulence needed for the DCF technique. The complication arises when there are more than one clearly distinguishable Gaussian components in the spectra for most of the region, like in the case of the GC. 
	
	This complexity of the GC environment makes it a challenging region to measure the magnetic field using the DCF method. Recent investigation by \citet{Chen2022} show that the line-width measurement which is used to constrain the turbulence along the LOS can be the major source of uncertainty or overestimation of the $B$-field when using the DCF method. The measured line-width can be a fairly reliable tracer of turbulence in regions where we can expect a single structure along the LOS, as in the case of filaments mostly at high Galactic latitudes. But when we look at the spectrum closer to the Galactic disk, it becomes evident that this approach becomes insufficient. A sample spectrum of the GC is shown in Fig. \ref{fig:sample_spectrum}. The spectrum indicates matter distributed in multiple layers along the LOS. Another factor to consider when using these data to measure the line-width is which component/components along the LOS contribute to the observed polarization. 
	
	\subsection{Spectral decomposition using \textsc{scousepy}}
	The decomposition of the CO spectrum was done using the Python implementation of the Semi-Automated multi-COmponent Universal Spectral-line fitting Engine \citep[\textsc{scousepy};][]{Henshaw2016,Henshaw2019}\footnote{\url{https://github.com/jdhenshaw/scousepy}}. This is a routine developed in Interactive Data Language (IDL) and later implemented in Python and is very useful to decompose complex spatial and velocity structures in spectral data cubes (hereafter: PPV cubes) in a systematic and efficient way. It is a multi-stage procedure and the important steps can be broken down as follows: (i) specify the region of interest based on position, velocity, or noise threshold; (ii) the routine then breaks the region into Spectral Averaging Areas (SAAs) based on the complexity of the spectrum in the region and extracts the spatially averaged spectrum from each; (iii) the extracted spectrum is manually fit interactively using \textsc{pyspeckit}\footnote{\url{https://github.com/pyspeckit/pyspeckit}}; (iv) the best fit solution of the extracted spectrum from each SAA is used to fit all the individual spectra within the SAA using a fully automated fitting procedure. The tolerance levels used in each step are described in \citet{Henshaw2016,Henshaw2019}. 
	
	In our application of \textsc{scousepy} on the data smoothed to the HAWC+ 216 \micron\ resolution, we masked all the pixels with peak signal below 1.5 K and used the width of the SAAs to be 10 pixels. We obtained 258 SAAs which were manually fit to automate the fitting for the 6362 individual spectra of the region. A similar procedure was performed to obtain the decomposed velocity components for the SCUPOL 850 \micron\ observation. At the end of the routine we obtain the number of velocity components in each pixel, their amplitude, shift, width, and a few additional useful statistics. A sample of the spectra decomposed using \textsc{scousepy} is shown at the bottom of Fig. \ref{fig:sample_spectrum}. We reject the components with signal-to-noise less than five and use only the robustly resolved components for further analysis.
	
	\begin{figure*}
		\centering
		\includegraphics[trim=0 2cm 0 0, scale=0.75]{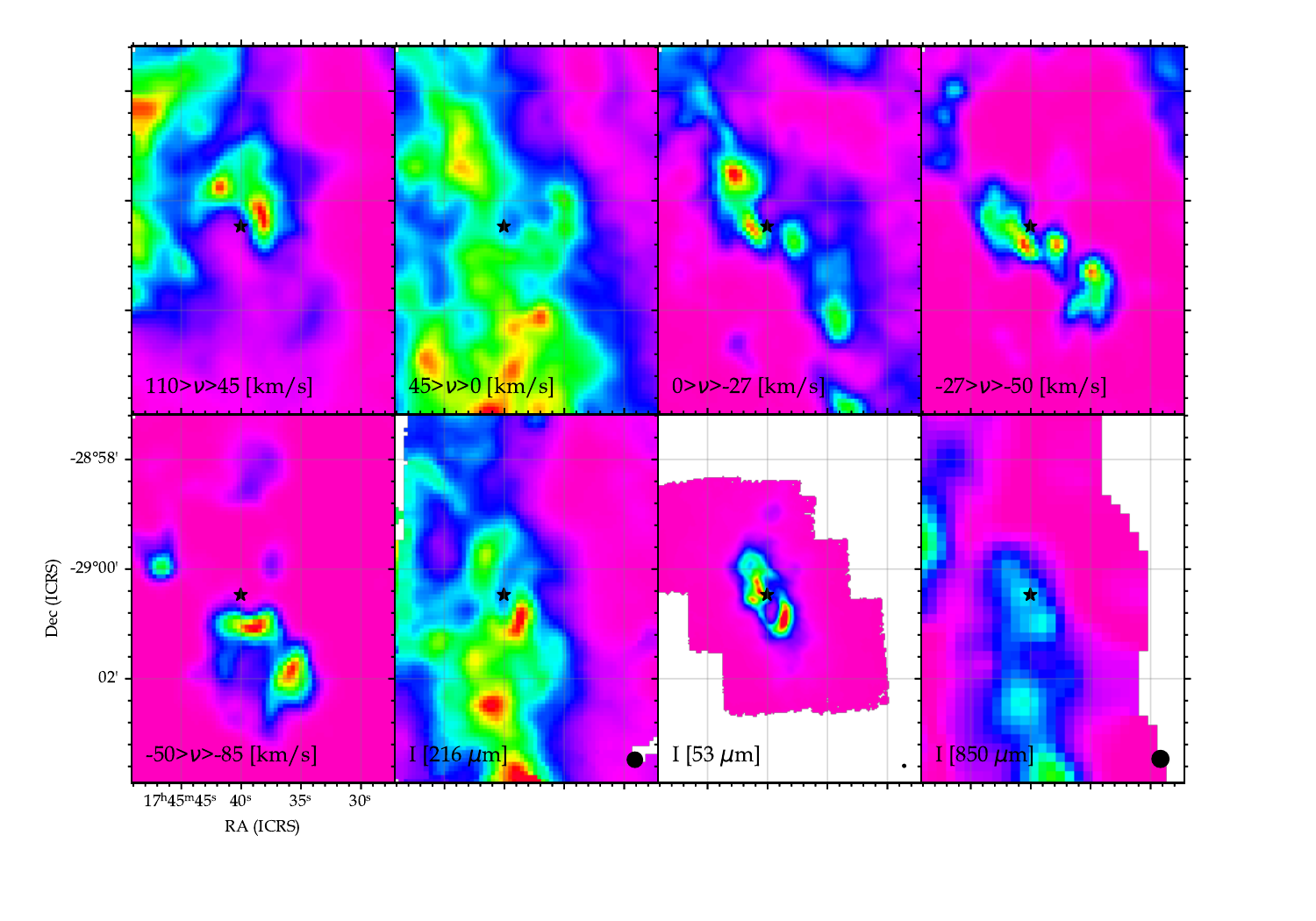}
		\caption{Moment zero maps of the PPV cube sub-slabs based on the clustering shown in Fig. \ref{fig:acorns}. The spectral cube was re-binned to match the SOFIA/HAWC+ 216 \micron\ observation. The Stokes I maps of HAWC+ 53 and 216 \micron\ and SCUPOL 850 \micron\ observations are also shown to compare the morphology of individual slabs with the morphology of intensity (Stokes I) observed in the polarization maps.}
		\label{fig:s216_components}
	\end{figure*}
	
	\begin{figure*}
		\centering
		\includegraphics[trim=0 1cm 0 0,scale=0.65]{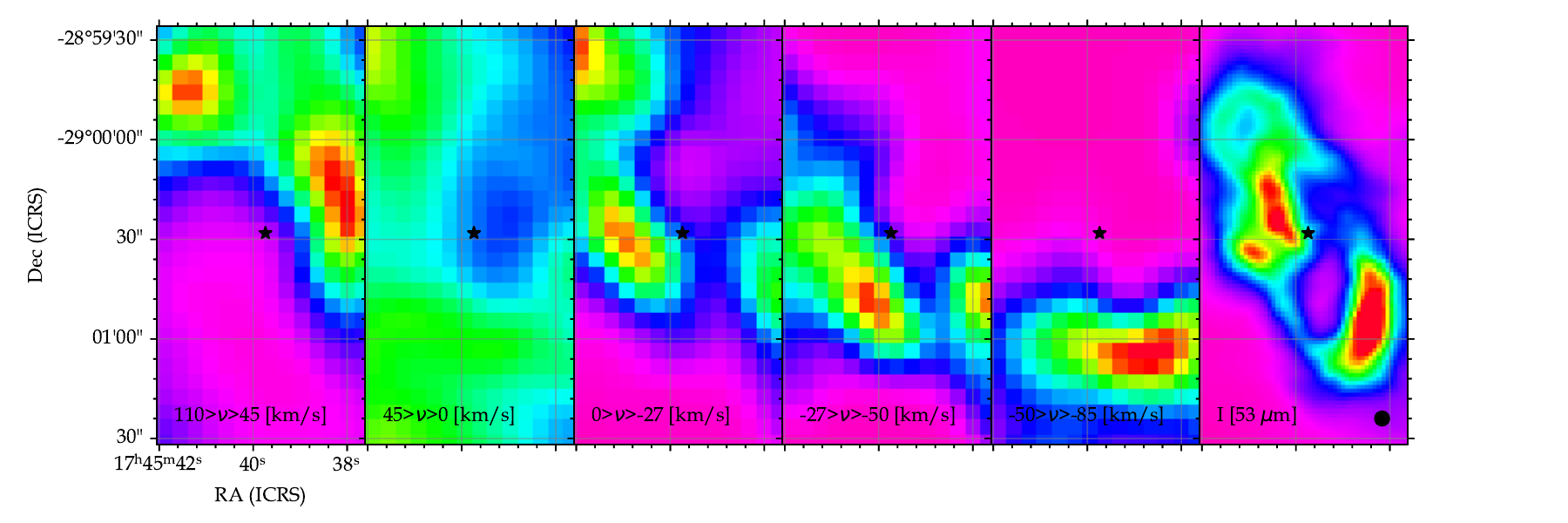}
		\caption{Same as Fig. \ref{fig:s216_components} but zoomed-in to focus on the region covered by the HAWC+ 53 \micron\ observation, which mostly contains the CND and the minispiral.}
		\label{fig:s53_components}
	\end{figure*}
	
	\subsection{Hierarchical Agglomerative Clustering using \textsc{acorns}}
	In order to characterise the decomposed velocity components from \textsc{scousepy} and understand its effects on the observed polarization, we have used a recently developed analysis tool based on hierarchical agglomerative clustering called \textsc{acorns} \citep[Agglomerative Clustering for ORganising Nested Structures;][]{Henshaw2019}\footnote{\url{https://github.com/jdhenshaw/acorns}}. The clustering procedure here begins with the most significant data point and a hierarchy is established by the merging of clusters based on user defined criteria. For PPV cubes like our data set, the clustering is performed using the position (x,y), intensity, and velocity. The criteria we used for the clustering analysis are as follows: (i) minimum peak intensity for the components to be considered was five times the typical root-mean-square (rms) value \citep{Henshaw2019} (ii) minimum size of clusters had to be 30 arcsec ($\sim1.5$ times the observation beam size) (iii) the difference in velocity between linked data points cannot be greater than 15 km s$^{-1}$. For visualization purposes, we chose trees from the forest which has a minimum of 4 leaves from the clustering procedure. This resulted in 8 prominent trees shown in Fig. \ref{fig:acorns}. From the figure it can be seen that though there seems to be a great number of velocity components towards the GC, they can still be grouped into meaningful distinct sub-structures. 
	
	The structures with negative centroid velocities centered at about $-60$, $-40$, and $-20$ km s$^{-1}$ are quite distinct and appear to be isolated structures without much overlap. These components are clearly seen in most of the spectra as shown in Fig. \ref{fig:sample_spectrum}. The positive velocity region however shows significant overlap in the observed features except for a distinction at around 30 km s$^{-1}$. This can also be seen in Fig. \ref{fig:sample_spectrum}, where the peaks at positive velocities blend each other more than the negative velocity peaks which are distinct. Based on these observed velocity features, we divided the initial spectral cube into sub-slabs with velocity ranges corresponding to the distinct trees from \textsc{acorns}. The moment zero maps of these slabs are shown in Fig. \ref{fig:s216_components}. We compared the morphology of the velocity features observed in the moment zero maps of each sub-slab with the intensity maps of the polarization observations in all three wavebands considered. The zoom-in version of the map for the region covered by the 53 \micron\ observation is shown in Fig. \ref{fig:s53_components}. 
	
	The key take away from this analysis is the difference in morphologies of the decomposed velocity structures and how they match the polarization observations at different wavelengths. Probing a region at different frequencies results in the observation of dust at different temperatures along the LOS, with shorter wavelengths originating from hot regions with high optical depth and longer wavelength optically thin emission originating from cooler deeper dust. When we are looking at regions that are known to have various independent structures along the LOS, we need to be careful with our measurement of line width to characterise turbulence as the chosen gaussian component might be probing a layer quite different from the source of our measured polarization. This effect is demonstrated in Fig. \ref{fig:s216_components} and Fig. \ref{fig:s53_components}. 
	
	When we look at the morphologies of the PPV sub-slabs and compare them with those in the Stokes I maps of HAWC+ 53 and 216 \micron\ observations, the 216 \micron\ observed intensity follows closely the morphology of the slab with the velocity range $0<v<45$ km s$^{-1}$. Though this velocity sub-slab seems to be the major contributor, there are still traces of features from other slabs in the intensity map of 216 \micron\ observation. This could indicate that the net observed emission at this wavelength might be an integrated effect from all the layers along the line of sight. However, this does not seem to be the case for the 53 \micron\ observation. When we compare its morphology with those of the moment maps from individual slaps, it becomes clear that most if not all of the observed emission at this wavelength originates from the negative velocity components of our decomposed spectra, and is clearly shown in Fig. \ref{fig:s53_components}. The 53 \micron\ observation is dominated by the emission from the CND and the minispiral. There does not seem to be any trace of the morphology observed from the positive velocity components in the Stokes I map of the 53 \micron\ observation. This is very important when we are using the spectra to constrain the turbulence in the region to measure the $B$-field. From this analysis, we propose that the right estimate of turbulence for the CND and the minispiral comes from the negative velocity components of the CO spectra and use only the velocity peaks in this range for the measurement of velocity dispersion for the 53 \micron\ observation. It is also interesting to note that the three negative velocity components shown in Fig. \ref{fig:s53_components} closely resemble the model for the velocity of the streamers around Sgr A$^*$ by \citet[][Fig. 21]{Zhao2009}. 
	
	\begin{figure}
		\centering
		\includegraphics[scale=0.5]{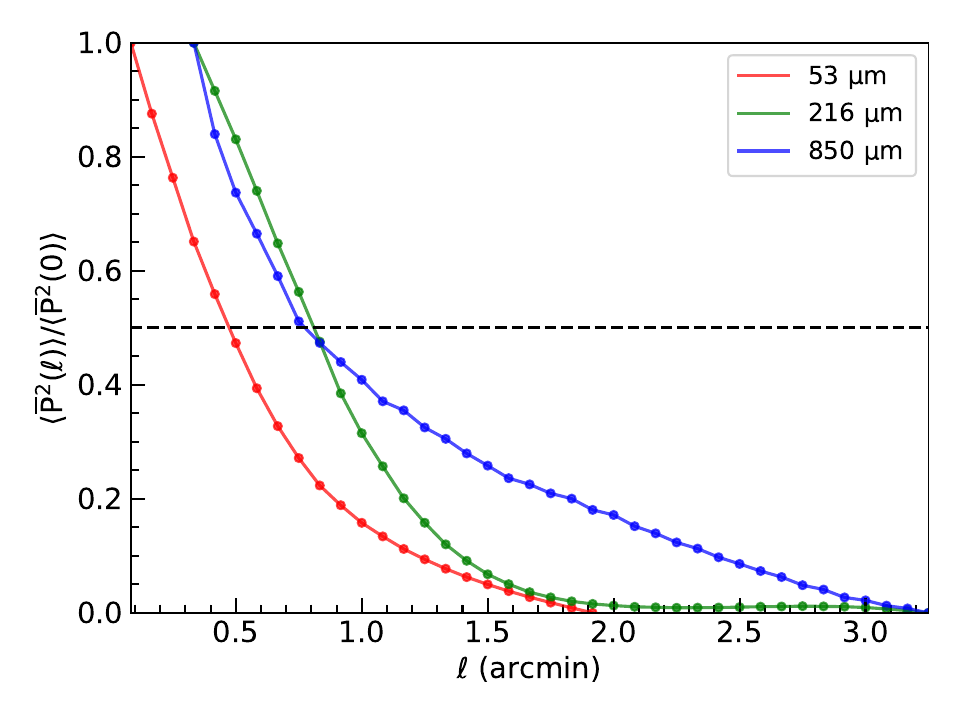}
		\caption{Normalized auto-correlation function calculated from polarized intensity for each of the observations. The value of $\Delta'$ is determined at half magnitude and corresponds to $0.47', 0.81',$ and $0.78'$ at 53, 216, and 850 \micron\ respectively.}
		\label{fig:Delta_Dash}
	\end{figure}
	
	\section{Measurements of the plane of sky B-field} \label{sec:B_field}
	We now measure the $B$-field strength using two modified DCF methods; (1) based on the angle dispersion structure function by \citet{Houde2009} and (2) the recently proposed Differential Measure Analysis technique by \citet{Lazarian2022}. The angle dispersion method takes into account the observed dispersion in the polarization due to the changes in the $B$-field orientation, but still assumes the turbulence to be isotropic. The Differential Measure Analysis method improves on this and accounts for the anisotropic nature of turbulence. Each of these methods is described in detail in the following sections.
	
	\subsection{$\mathbf{B_{{}_{\mathrm{POS}}}}$ from Angle Dispersion Function} 
	\citet{Houde2009} improved the original DCF technique by using the structure function of polarization angles to calculate the polarization angle dispersion such that the dispersion due to the change in field orientation is also accounted for along with the dispersion due to turbulence. Following \citet{Houde2009}, Equation (\ref{eq:1}) can be rewritten as;
	\begin{equation}\label{eq:B_Houde}
		B_{{}_{\mathrm{POS}}} = \sqrt{4\pi\rho} \sigma_v \left[ \frac{\langle B_t^2 \rangle}{\langle B_0^2 \rangle} \right]^{-1/2},
	\end{equation}
	where $\langle B_t^2\rangle/\langle B_0^2\rangle$ is the square of the turbulent to large-scale magnetic field strength ratio and can be approximated as the dispersion in the polarization vectors ($\sigma_{\phi}=[\langle B_t^2\rangle/\langle B_0^2\rangle]^{1/2}$). This quantity can be determined using the two-point dispersion function proposed by \citet{Houde2009}, modelled as a superposition of the large-scale field structure and the small-scale turbulence. This method takes care of the spatial changes in the $B$-field morphology and also incorporates radiative-transfer effects and smoothing by the telescope beam. If $\Delta\phi(\ell)$ is the difference in polarization angle between two vectors separated by an angle $\ell$ on the sky, then the dispersion function is defined as;
	\begin{equation}
		1 - \langle \mathrm{cos}\left[\Delta\phi(\ell)\right]\rangle = \frac{1 - e^{-\ell^2/2(\delta^2+2W^2)}}{1+\mathcal{N}\left[\frac{\langle B_t^2\rangle}{\langle B_0^2\rangle}\right]^{-1}} + a_2\ell^2,
	\end{equation}
	where $\delta$ is the turbulence correlation length, $W$ is the observations beam ``radius'' (assuming a Gaussian beam) defined as the distance from the beam axis where the intensity drops to $1/e^{1/2}$ and the corresponding values are 2.05$''$, 7.71$''$, and 8.48$''$ for the 53, 216, and 850 \micron\ observations respectively, $\mathcal{N} = \Delta'(\delta^2+2W^2)/\sqrt{2\pi}\delta^3$ is the number of turbulent cells along the line of sight, and $\Delta'$ is the effective depth of the cloud. The small-scale $B$-field contribution to the observed dispersion is quantified by the first term in the equation and the large-scale contribution is described by the second term.
	
	We use this angle dispersion structure function to determine $\sigma_{\phi}$ on a pixel-by-pixel basis for each of the observations we have chosen. Before we apply the dispersion function to the whole image, we need to determine the values of $\Delta'$. The depth of the cloud is determined using the width of the auto-correlation function of the polarized intensity ($P=\sqrt{Q^2+U^2}$) given by;
	\begin{equation}
		\langle\overline{P}^2(\ell)\rangle\equiv\langle\overline{P}(r)\overline{P}(r+\ell)\rangle,
	\end{equation}
	as described in \citet{Houde2009}. Based on the assumption that the cloud has similar characteristics across and along its depth, this is a reasonable approximation of the effective depth of the cloud even though it is a function of $\ell$ along the cloud surface. We calculated the normalized auto-correlation function for each of the observations and the corresponding plot is shown in Fig. \ref{fig:Delta_Dash}. The estimated values of $\Delta'$ at 53, 216, and 850 \micron\ are $0.47'$, $0.81'$, and $0.78'$ respectively.

	In order to estimate the dispersion function locally around each pixel, it is important to choose a kernel size ($w$) in pixel radius over which $1 - \langle \mathrm{cos}\left[\Delta\phi(\ell)\right]\rangle$ is estimated for each pair of polarization angles. This symmetric two-dimensional normalized circular kernel ensures no preferential direction is chosen when estimating the dispersion function. The size has to be greater than the observation beam size and the turbulence correlation length so that we have enough vectors within the kernel to calculate the dispersion function with greater accuracy. We used the method described by \cite{Guerra2021} to determine the optimal kernel size for each observation. As each of our polarization maps probes a different physical scale with quite different resolutions, the corresponding kernel size could also be different. We fit the dispersion function to every pixel for a range of kernel sizes from $w  = 5-15$ pixels and calculated the Spearman's correlation coefficient ($\rho_{\mathrm{sp}}$) for each kernel size. The best size was chosen based on the highest median value of estimated $\rho_{\mathrm{sp}}$ and was found to be 9, 7, and 13 pixels for the 53, 216, and 850 \micron\ observations, respectively. 
	
	\begin{table}
		\caption{DCF method parameters.}
		\resizebox{\columnwidth}{!}{%
			\begin{tabular}{|c|c|c|c}
				\hline
				Parameter                                                    & 53 \micron\        & 216 \micron\            & 850 \micron             \\ \hline  \hline
				beam size (arcsec)                                            & $4.84$             & $18.2$                  & $20$                    \\
				pixel size (arcsec)                                          & $2.42$             & $4.55$                  & $10$                    \\
				$\Delta'$ (arcmin)                                           & $0.47$             & $0.81$                  & $0.78$                  \\
				$w$ (pixels)                                                 & 9                  & 7                       & 13                      \\
				$\delta_{\mathrm{fit}}$ (arcsec)                             & $21.04\pm0.49$ & $22.79\pm2.6$       & -                       \\
				$[\langle B_t^2\rangle/\langle B_0^2\rangle]_{\mathrm{fit}}$ & $0.16\pm0.003$ & $0.016\pm0.002$     & -                       \\
				$[\langle B_t^2\rangle/\langle B_0^2\rangle]_{\mathrm{mean}}$ & $0.29\pm0.5$ & $0.017\pm0.02$                 & $0.38\pm0.09$                   \\
				\multirow{2}{*}{$\sigma_v$ (km/s)}                           & $5.29\pm2.16$                & \multirow{2}{*}{$8.68\pm1.95$} & \multirow{2}{*}{$9.75\pm1.83$} \\
				& $(8.74\pm1.66)^*$        &                         &                         \\
				$\sigma_{\phi}$                                         & $0.42\pm0.34$             & $0.12\pm0.06$                  & $0.61\pm0.08$                  \\
				$n_{{}_{\mathrm{H}}}^{\mathrm{mean}}$ (cm$^{-3}$)            & $10^{4.2}$         & $10^{3.93}$       & $10^{3.94}$       \\
				\multirow{2}{*}{$B_{{}_{\mathrm{POS}}}^{\mathrm{mean}}$ (mG)} & $2.03\pm1.8$ & \multirow{2}{*}{$6.76\pm3.49$} & \multirow{2}{*}{$1.43\pm0.57$} \\ 
				& ($3.28\pm2.37$)$^*$ &                         &                         \\\hline
			\end{tabular}%
		}
		\label{tab:dispersion_parameters}
		\footnotesize{$^*$ estimates from all the velocity components. All the quoted errors are the 1$\sigma$ standard deviation values.}
	\end{table}
	
	Using the above described values of $\Delta'$ and $w$, we fit the dispersion function for each polarization observation to determine the parameters $\delta$, $a_2$, and $\langle B_t^2\rangle/\langle B_0^2\rangle$ and the corresponding fit parameters are shown in Table. \ref{tab:dispersion_parameters}. However, we found most of the estimated $\delta$ values ($\sim 50$\%) for the 216 and 850 \micron\ observations to be $\delta<W\sqrt{2}$ which is the correlated beam size of the observations. This indicates that the resolution is not good enough to resolve the local gas turbulence and might lead to incorrect estimates of the $B$-field. Thus we tried another approach presented in \cite{Guerra2021} where we fix the value of $\delta$ based on the fit of a single dispersion function for the whole region (global $\delta$). The best fit $\delta$ was found to be $22.79''\pm2.6''$ for the 216 \micron\ observation. We used the same value for the 850 \micron\ observation as they have similar beam sizes ($18.2''$ and $20''$ respectively) and due to the lower data quality of the 850 \micron\ observation.  
	
	\begin{figure}
		\centering
		\includegraphics[scale=0.5]{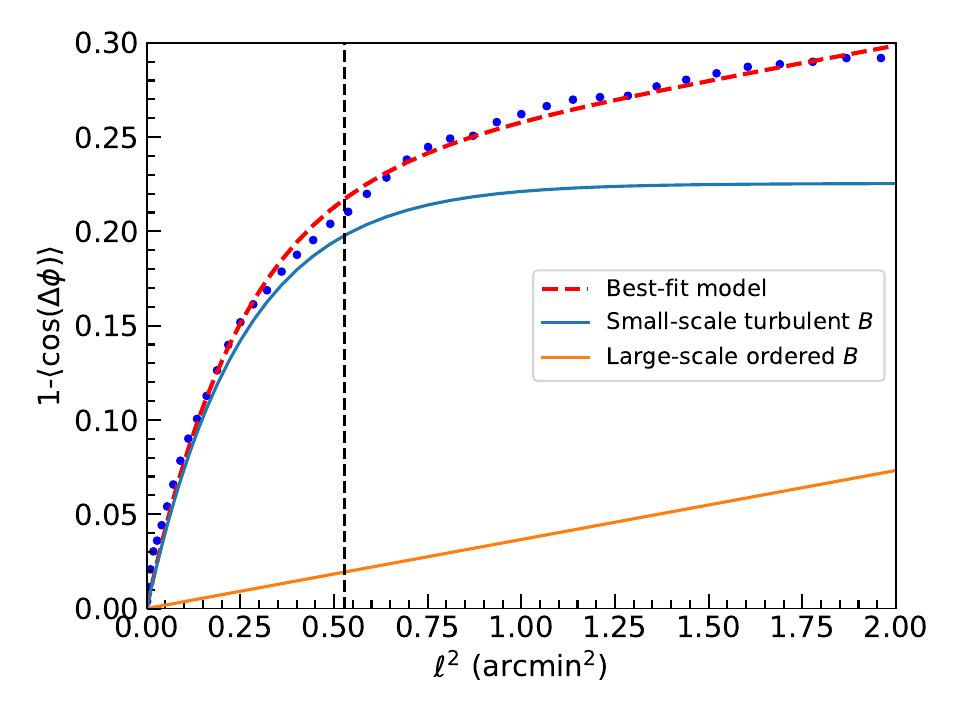}
		\includegraphics[scale=0.5]{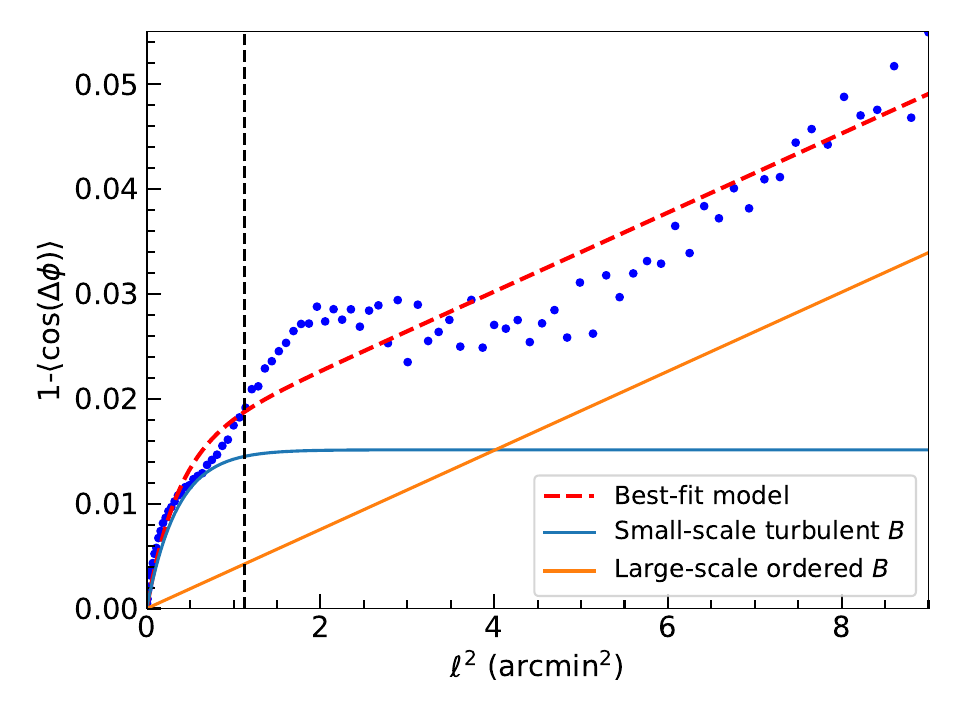}
		\caption{Dispersion function fit to the entire field of view for the 53 \micron\ (top) and the 216 \micron\ (bottom) observations. The blue dots are the estimates from the observations and the dashed red line is the best fit model for each case. The first term of the dispersion function which gives an estimate of the small-scale turbulence component is shown as the cyan line and the large-scale mean field component is shown by the orange line. The size of the kernel chosen ($w=9$ and 7 for 53 and 216 \micron\ observations, respectively) for each of the maps is shown by the black dotted line in both plots.}
		\label{fig:Houde_component}
	\end{figure}
	
	\begin{figure*}
		\centering
		\includegraphics[scale=0.5]{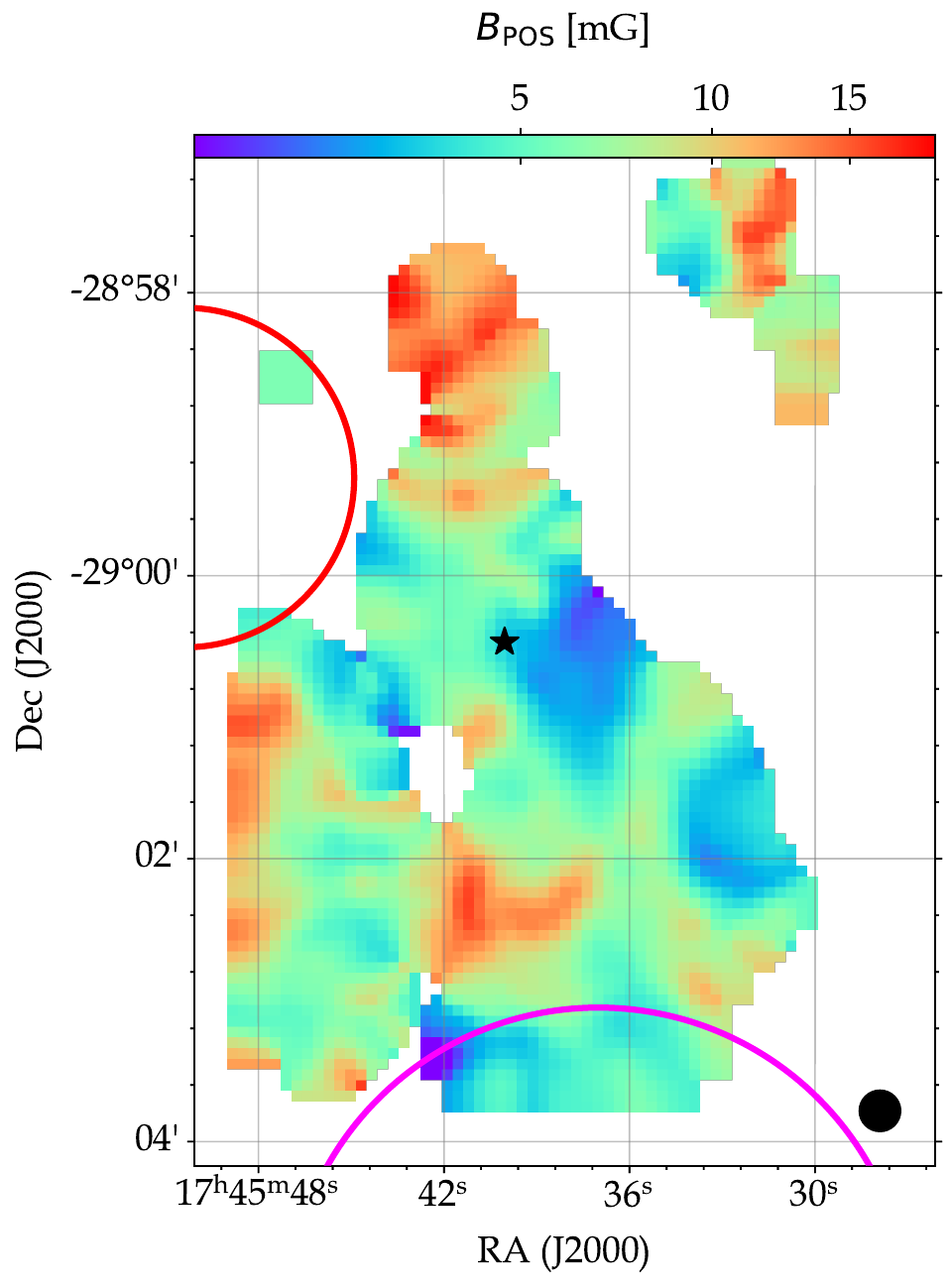}
		\hspace{1cm}
		\includegraphics[scale=0.5]{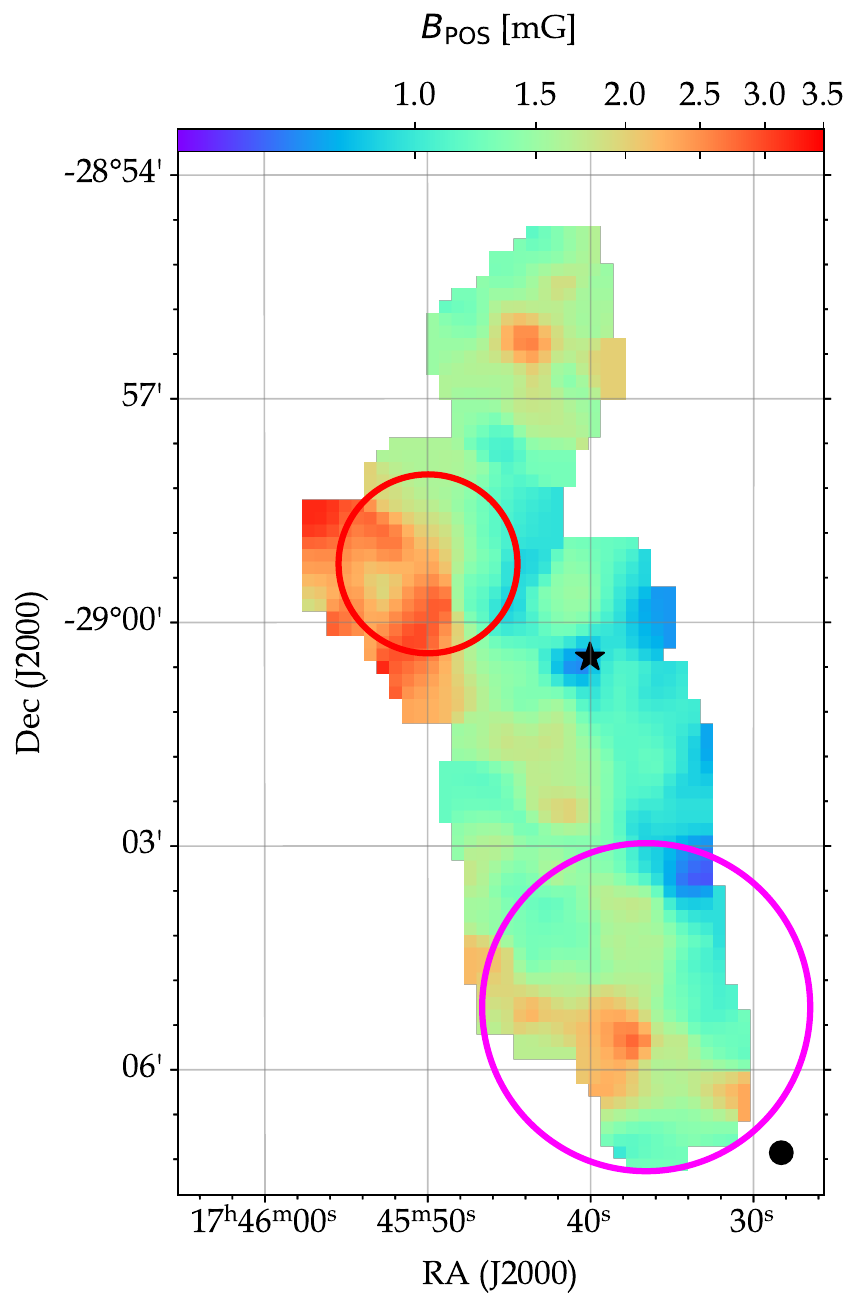}
		\caption{Maps of the $B_{\mathrm{POS}}$ estimated from HAWC+ 216 \micron\ (left) and SCUPOL 850 \micron\ (right) observations using the DCF method. The location of the 50 km s$^{-1}$ and 20 km s$^{-1}$ clouds are shown as the red and magenta circles, respectively. The beam sizes of the observations are shown at the bottom of each map.}
		\label{fig:B_field_s216_j850}
	\end{figure*}
	
	We tried both the approaches of a fixed $\delta$ and using $\delta$ as a free parameter for the 53 \micron\ observation. When $\delta$ is a free parameter, as in the previous cases about 40\% of the pixels still showed $\delta<W \sqrt{2}$. Hence we chose to fix the $\delta$ value to $21.04''\pm0.49''$ obtained from a single dispersion function fit to the whole image. There were some pixels with large error bars due to missing data in the surrounding region within the kernel size ($w=9$). In these cases we increased the kernel size to 11 to get better estimates of the parameters. By using this method we were able to get reasonable values of $\langle B_t^2\rangle/\langle B_0^2\rangle$ for most of the pixels of the observation ($\sim90\%$). 
	
	The plots of the single dispersion function fit to the 53 \micron\ and 216 \micron\ data are shown in Fig. \ref{fig:Houde_component}. The small-scale turbulent (first term of the dispersion function) and the large-scale mean field (last term) components are also shown. It can be seen from the plots that the turbulent component dominates the dispersion when $\ell\lesssim44''$ and $\ell\lesssim1'$ in 53 and 216 \micron\ observations, respectively. Beyond this size there is a significant contribution from the large-scale component to the observed dispersion. These sizes are also around the same as the kernel sizes chosen for each observation. 
	
	The best fit values of $\langle B_t^2\rangle/\langle B_0^2\rangle$ from each observation was used to estimate $\sigma_{\phi}$ using the approximation $\sigma_{\phi}\simeq[\langle B_t^2\rangle/\langle B_0^2\rangle]^{1/2}$. The mean value of $\langle B_t^2\rangle/\langle B_0^2\rangle$ was found to be $0.29\pm0.5$, $0.017\pm0.02$, and $0.38\pm0.09$ for 53, 216 and 850 \micron\ observations respectively. The 216 \micron\ observation shows the least dispersion which reflects the rather uniform polarization vectors observed in Fig. \ref{fig:pol_maps} \& \ref{fig:s216-pilot}. This might be a result of the majority of the dust emission at this wavelength originating from a region of strong magnetic field, and will be discussed in the following sections.  
	
	\begin{figure*}
		\includegraphics[trim={1cm 2cm 0 0}, scale=0.48]{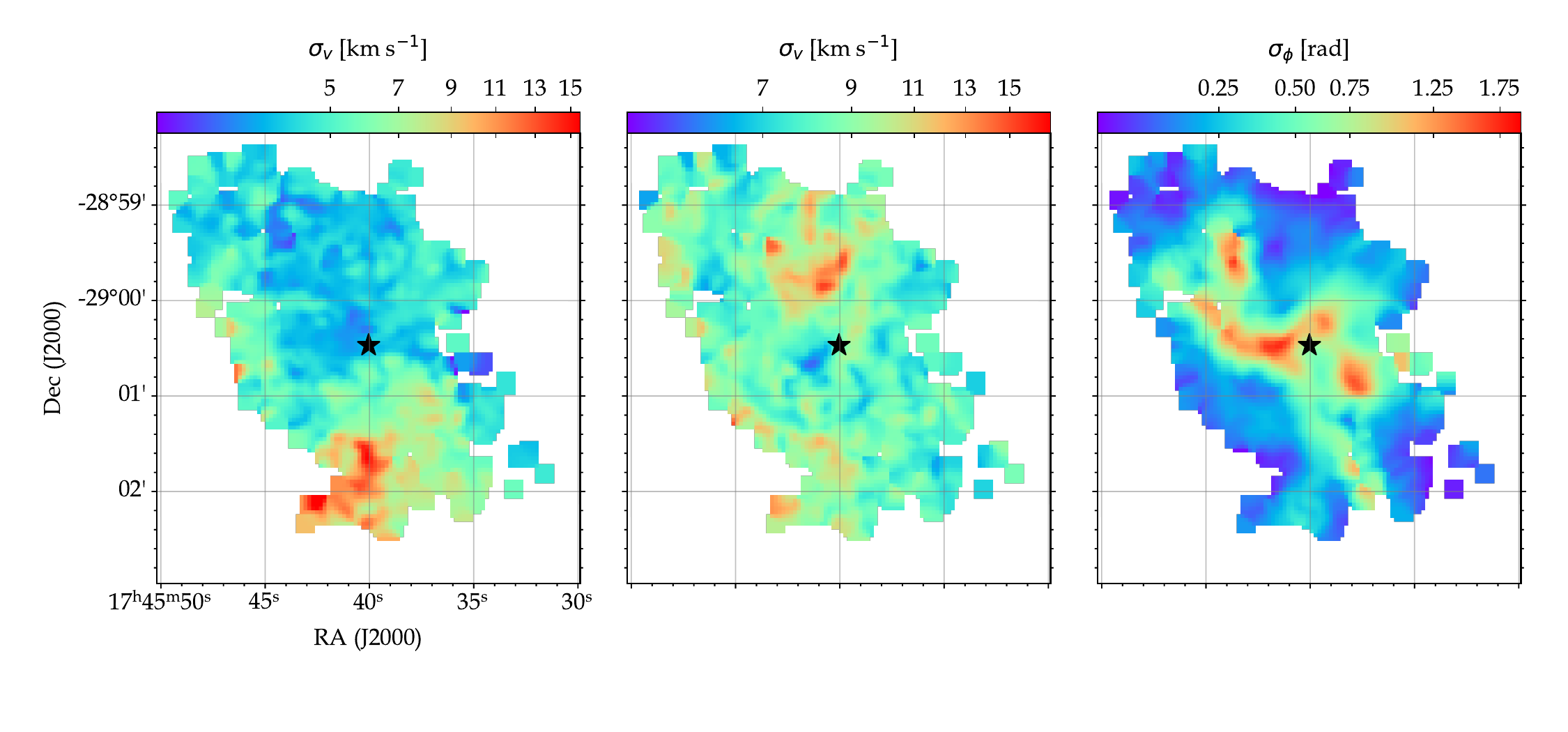}
		\caption{Maps of $\sigma_v$ from the negative velocity components (left), all velocity components (middle), and $\sigma_{\phi}$ (right) for the 53 \micron\ HAWC+ observation used to estimate the magnetic field of the region from DCF method.}
		\label{fig:s53_V_phi}
	\end{figure*}
	
	\begin{figure*}
		\centering
		\includegraphics[trim={1cm 2cm 0 0},scale=0.48]{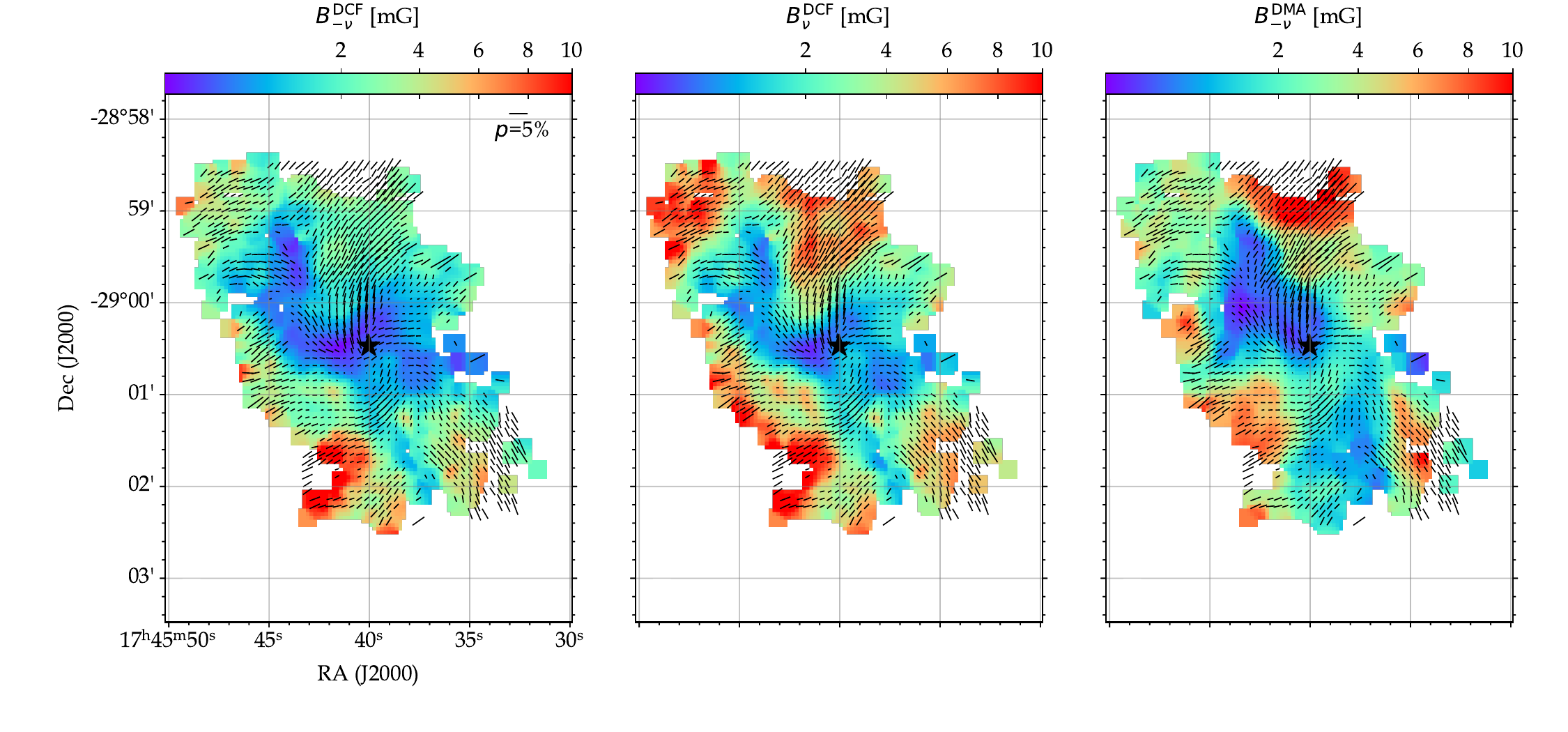}
		\caption{$B_{{}_{\mathrm{POS}}}$ for the 53 \micron\ observation estimated using the DCF method from the negative velocity components is shown on the left and from all the velocity components is shown in the middle. The map on the right shows the $B$-field estimate from the DMA method. The mean field from each map is 2.03$\pm$1.8 mG, 3.28$\pm$2.37 mG, and 2.79$\pm$2.25 mG respectively. The HAWC+ polarization vectors are overlaid on all the maps with its scale shown at the top of the map on the left.}
		\label{fig:B_field_s53}
	\end{figure*}
	
	\begin{figure*}
		\centering
		\includegraphics[scale=0.35]{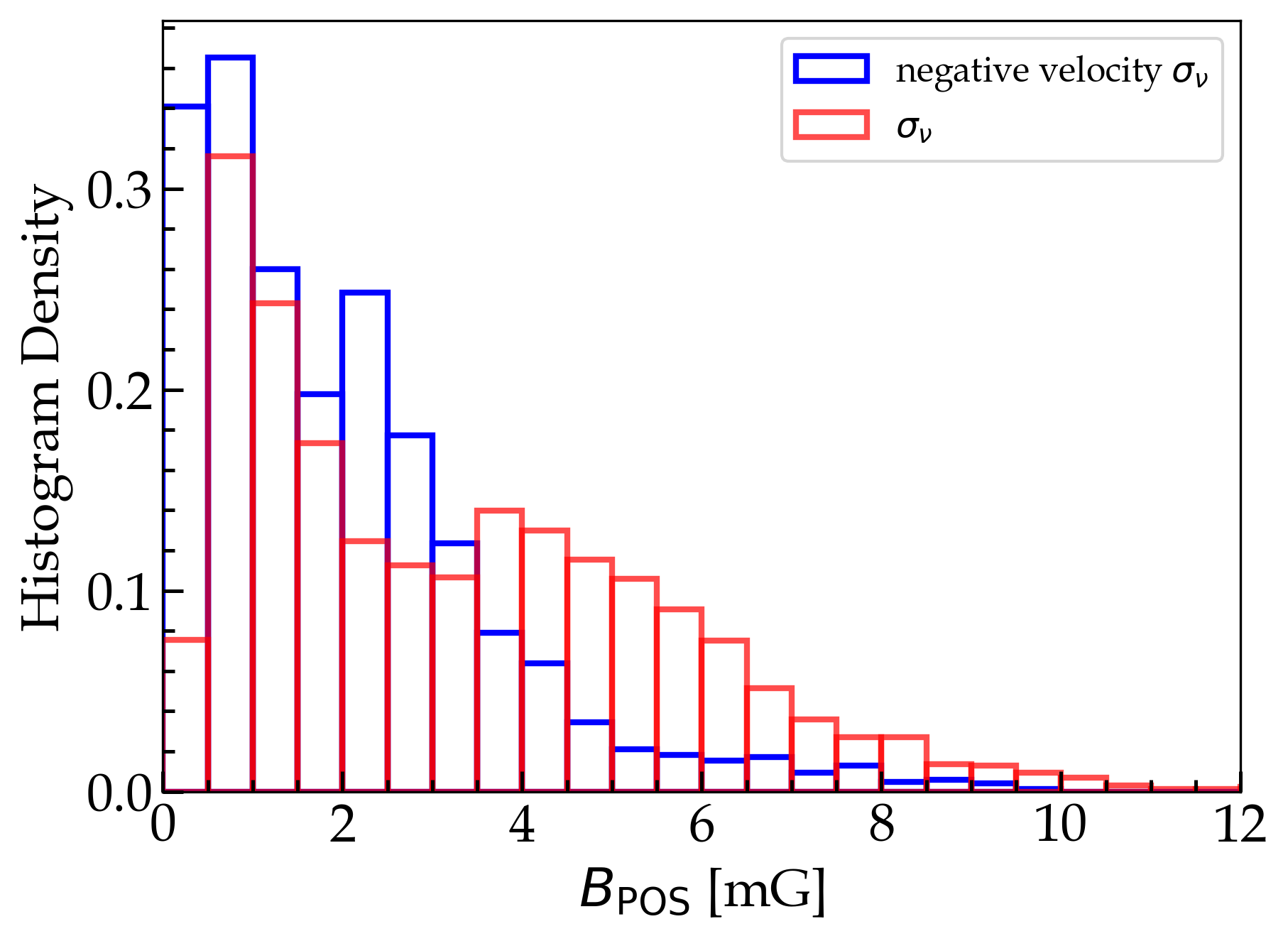}
		\includegraphics[scale=0.35]{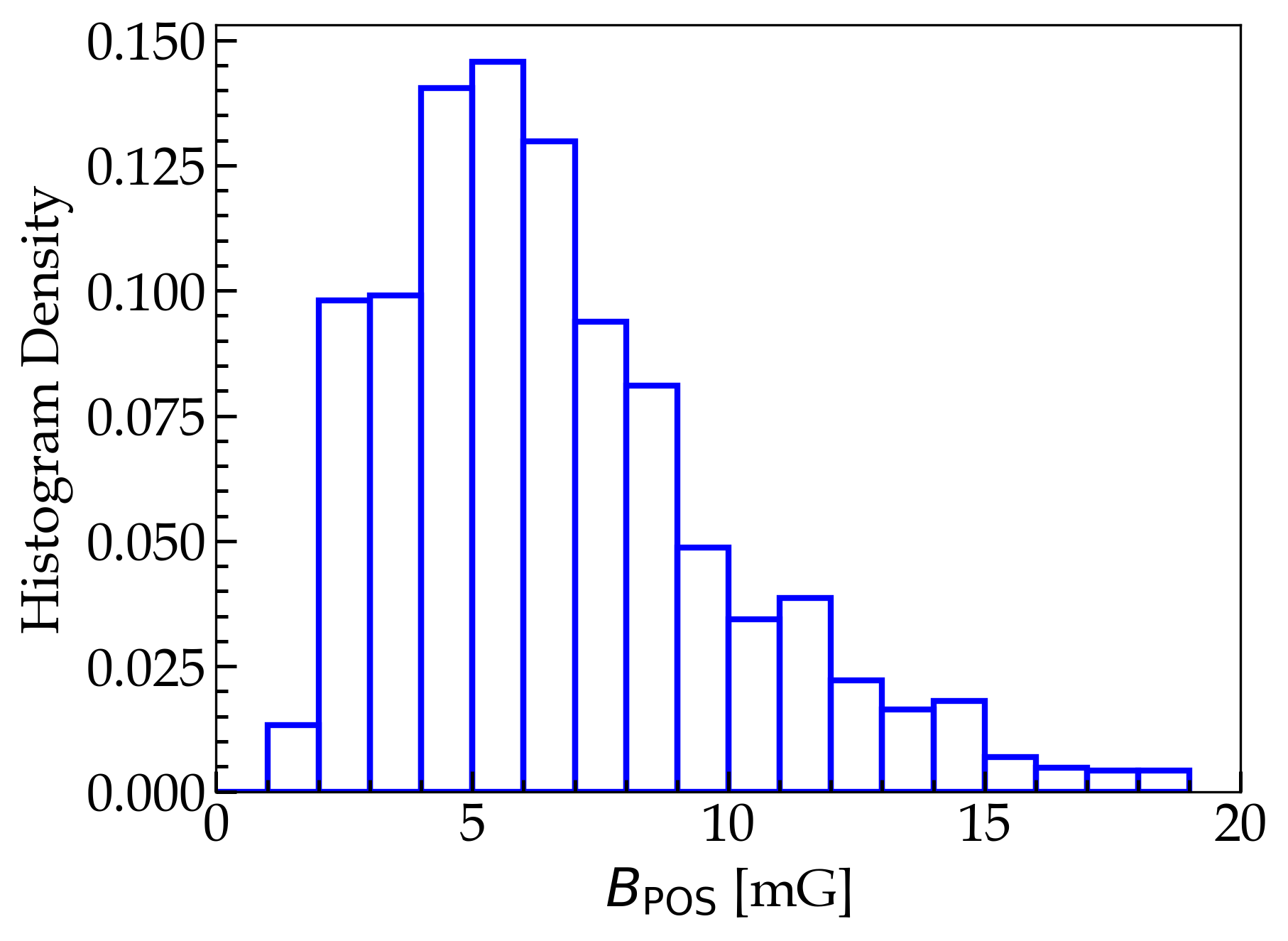}
		\includegraphics[scale=0.35]{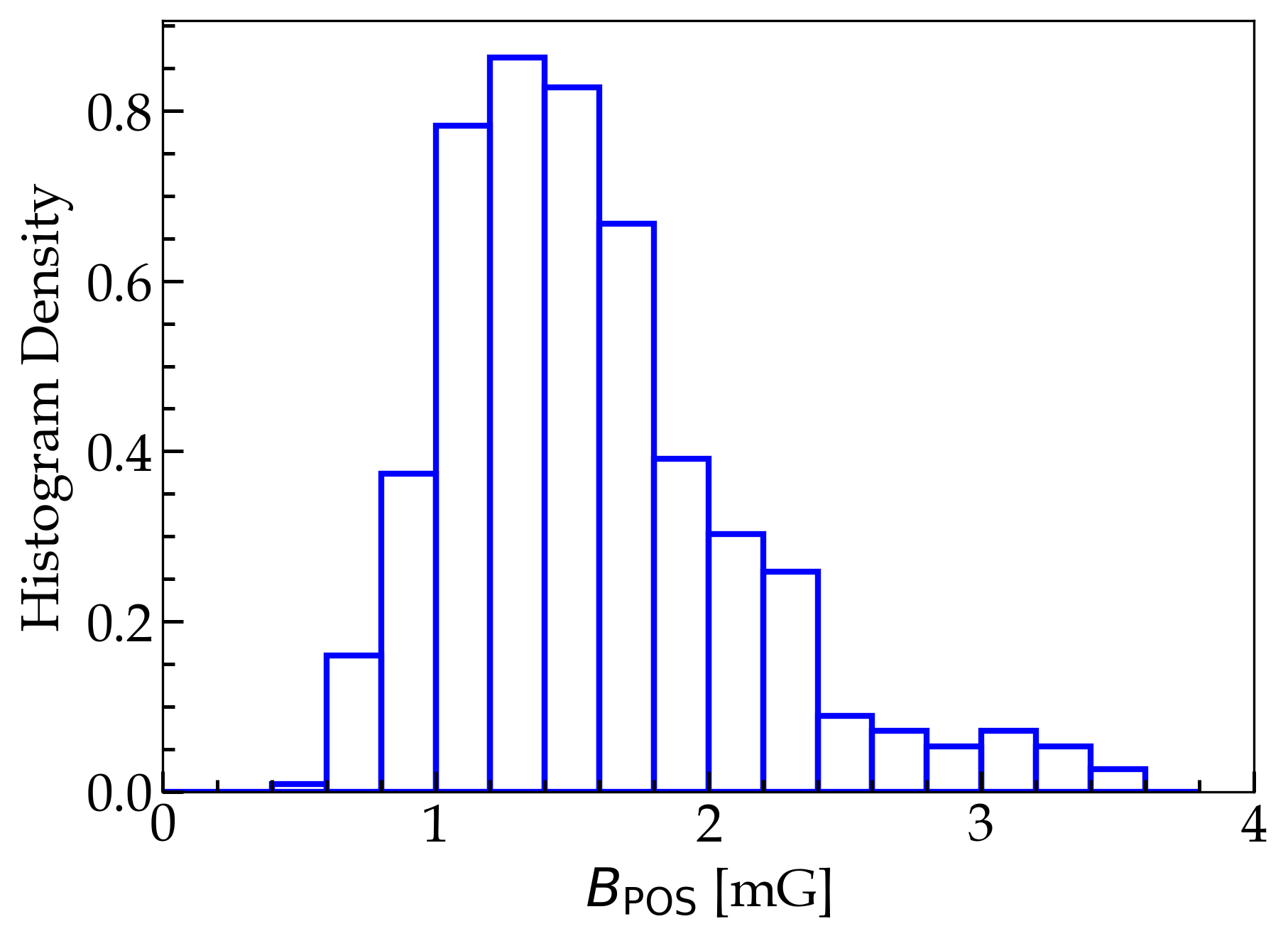}
		\caption{Distribution histograms of the estimated $B_{{}_{\mathrm{POS}}}$ from the 53 (left), 216 (middle), and 850 (right) \micron\ observations.}
		\label{fig:B_hist}
	\end{figure*}
	
	Now, we measure the $B_{{}_{\mathrm{POS}}}$ using the modified DCF method described in Equation \ref{eq:B_Houde}. The mass density was determined from the gas volume density ($n_{{}_{\mathrm{H}}}$) using the relation $\rho=\mu m_{{}_{\mathrm{H}}}n_{{}_{\mathrm{H}}}$, where $m_{{}_{\mathrm{H}}}$ is the mass of hydrogen and $\mu=2.8$ is the mean molecular weight per unit mass of hydrogen \citep{Sears1964,Cox2000,Kauffmann2008,Lodders2021}. The volume densities were determined using the depth of the clouds estimated from the auto-correlation function and the column density measurements from \textit{Herschel} observations. The resulting values match well with the earlier predictions using molecular line observations \citep[$n_{{}_{\mathrm{H}}}\simeq10^{4.1}$ cm$^{-3}$;][]{Oka2011}. More details about the estimation of temperature and gas column density of the region can be found in \citet{Akshaya2023}. 
	
	The velocity dispersion described in Section \ref{sec:Velocity_Dispersion} was corrected for the contribution from molecular thermal motion to extract only the dispersion due to turbulence using the relation $\sigma_v^2=\sigma_{v0}^2-k_\mathrm{B}T_{\mathrm{gas}}/m_{\mathrm{CO}}$, where $\sigma_{v0}$ is the dispersion measured from the molecular spectroscopic data, $k_{\mathrm{B}}$ is the Boltzmann constant, $T_{\mathrm{gas}}$ is the temperature of gas in the region, and $m_{\mathrm{CO}}$ is the mass of CO. We use the same approximation of $T_{\mathrm{gas}}=T_{\mathrm{dust}}$ as in our previous study. We have also ignored the pixels where the best fit values of $\langle B_t^2\rangle/\langle B_0^2\rangle<0.001$ as these are found mostly at the boundaries of the polarization observations and might be a result of poor fitting due to insufficient data. 
	
	The $B_{{}_{\mathrm{POS}}}$ of the 216 and 850 \micron\ observations were estimated using the mean $\sigma_v$ from all the resolved components shown in Fig. \ref{fig:acorns}. The resultant maps of the magnetic field are shown in Fig. \ref{fig:B_field_s216_j850} and have a mean value of 6.76$\pm$3.49 mG and 1.43$\pm$0.57 mG for 216 and 850 \micron\ observations, respectively. The high magnetic field estimate of the 216 \micron\ observation is due to the very low dispersion observed in the polarization measurement at this wavelength, as can be seen from the rather uniform polarization vectors in Fig. \ref{fig:pol_maps} \& \ref{fig:s216-pilot}. A histogram of the distribution of the magnetic field for each wavelength is shown in Fig. \ref{fig:B_hist}. It is important to note that these estimates are from the DCF method without any correction factor applied to the equation, hence the mean $B$-field strength should be treated as an upper limit of the field measured at these wavelengths.
	
	\subsubsection{B$_{{}_{\mathrm{POS}}}$ from resolved velocity components}
	We have used a different approach to measure the $B$-field of the 53 \micron\ observation. As discussed in Section. \ref{sec:Velocity_Dispersion}, the morphology of the observed emission at the 53 \micron\ wavelength matches well with the integrated map of the velocity components from the negative velocity region of the CO ($J=3\rightarrow2$) spectrum. Not taking this into account while using the $\sigma_v$ value will lead to an overestimation of the $B$-field. To understand the importance of resolving and isolating the velocity components in such complex regions we estimate the $B$-field from two values of $\sigma_v$. In one case we consider the mean $\sigma_v$ only from the negative velocity Gaussian components of the spectrum while in the other case, we take the mean value from all the resolved components. The maps of $\sigma_v$ for the two cases and the $\sigma_{\phi}$ for this region are shown in Fig. \ref{fig:s53_V_phi}. Similar to the previous case, we have ignored the pixels with $\langle B_t^2\rangle/\langle B_0^2\rangle<0.01$ at the boundaries of the polarization data. The resulting $B_{{}_{\mathrm{POS}}}$ maps are shown in Fig. \ref{fig:B_field_s53}. There is a clear difference in the estimated strength especially in the regions away from the central source Sgr A$^*$. The mean field of the region estimated from the negative velocity components is $B_{-v}^{\mathrm{DCF}}=2.03\pm1.8$ mG and the estimate by considering all the components is $B_v^{\mathrm{DCF}}=3.28\pm2.37$ mG. The distribution of the field from both cases is shown in Fig. \ref{fig:B_hist}. The field from all the components shows two peaks in the distribution, at around 1 mG and 4 mG.

	\begin{figure}
		\centering
		\includegraphics[scale=0.5]{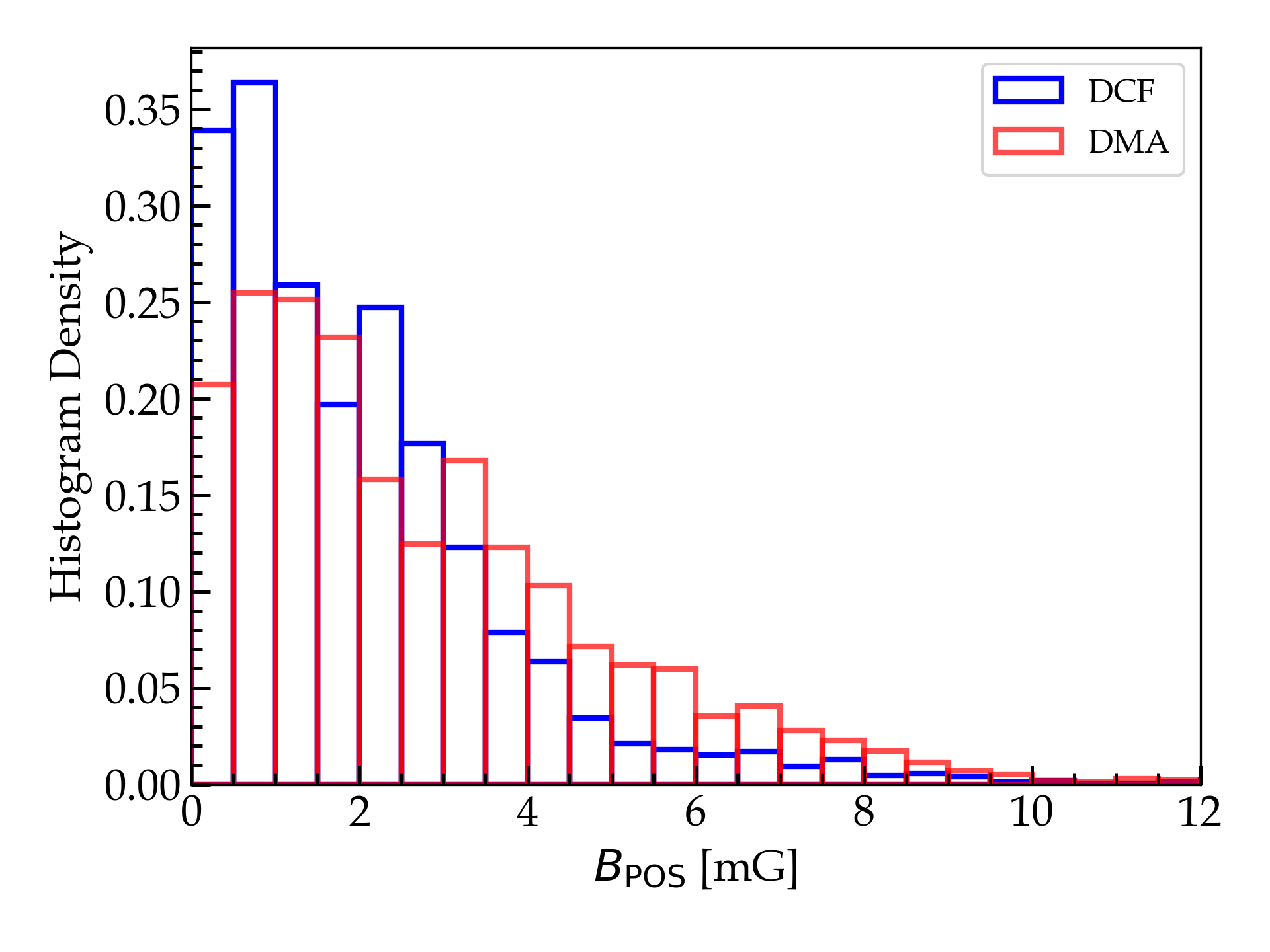}
		\caption{Histogram comparing the $B_{{}_{\mathrm{POS}}}$ measured from DMA and DCF methods for the 53 \micron\ observation, using correction factor $f=1$.}
		\label{fig:s53_DMA_hist}
	\end{figure}
	
	\begin{figure}
		\centering
		\includegraphics[scale=0.5]{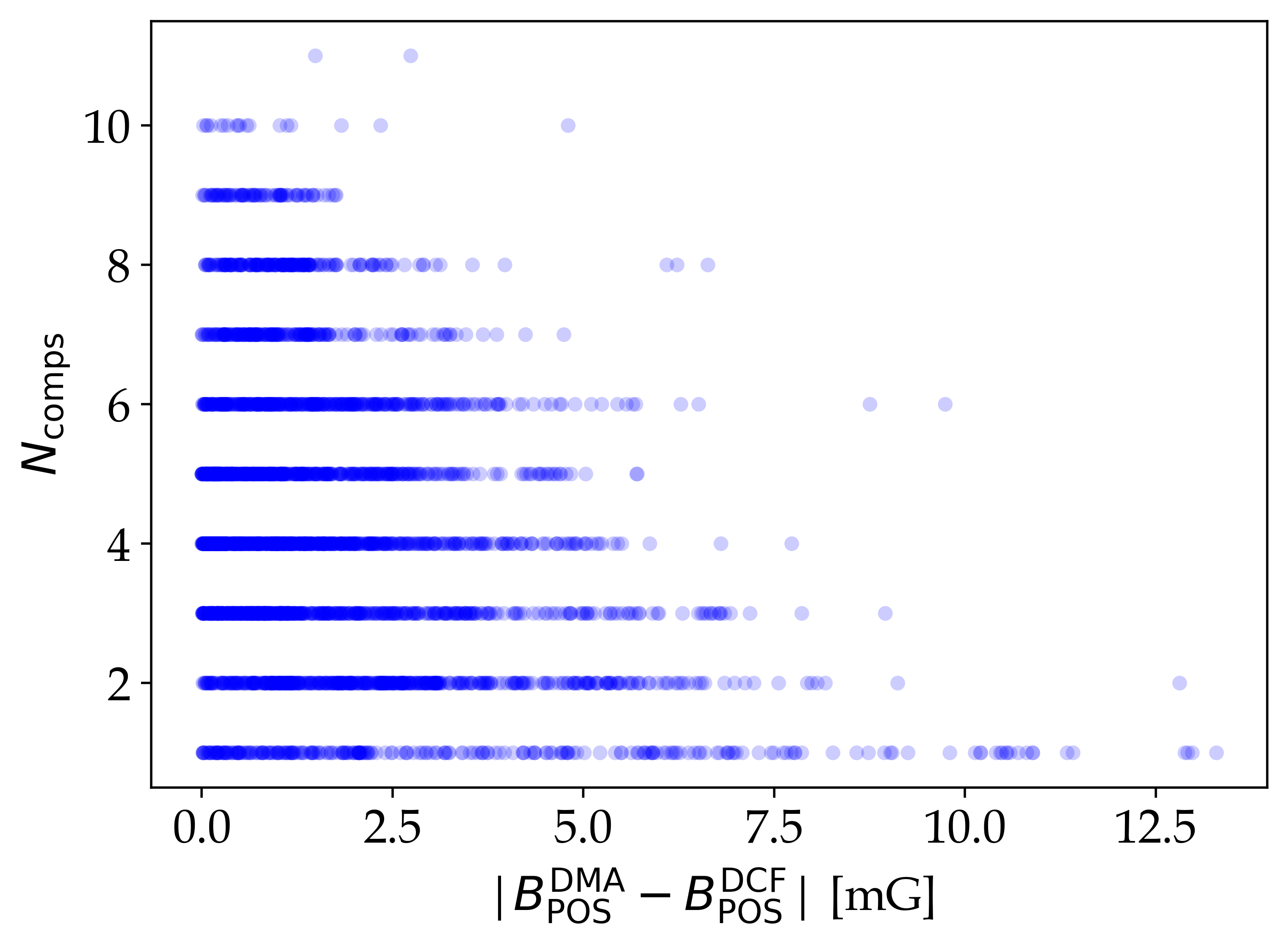}
		\includegraphics[scale=0.5]{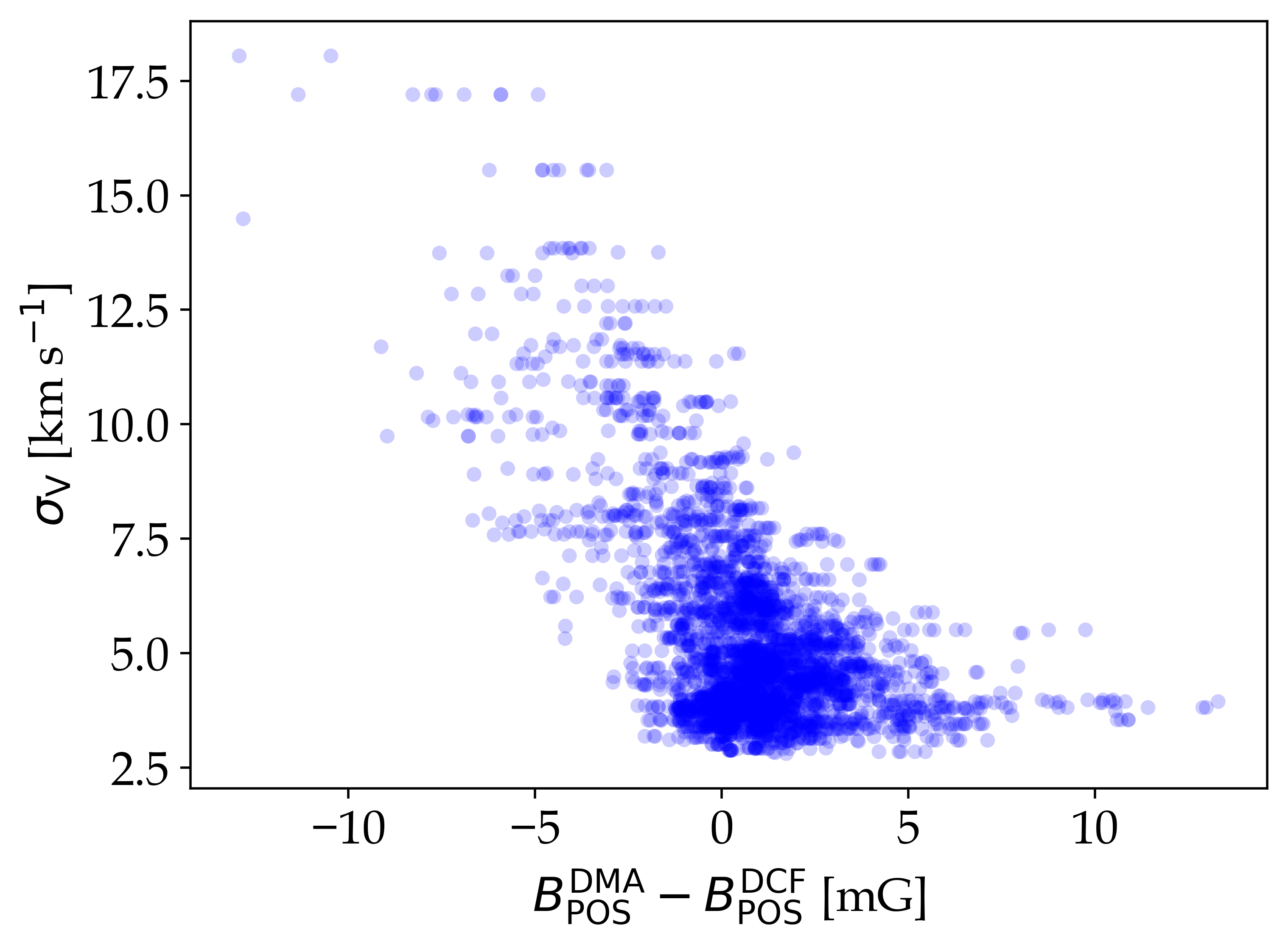}
		\caption{Correlation between the estimated difference in the $B_{{}_{\mathrm{POS}}}$ measured from DMA and DCF methods with the number of components resolved along the line of sight (top) and the measured velocity dispersion (bottom).}
		\label{fig:s53_Bdiff_dcf_dma}
	\end{figure}
	
	\subsection{$\mathbf{B_{{}_{\mathrm{POS}}}}$ from Differential Measure Analysis}
	We also measure the $B$-field strength using another recent modification of the original DCF method, called the Differential Measure Analysis (DMA) technique proposed by \citet{Lazarian2022}. The observed dispersion in the polarization is an effect of the competition between the turbulence-induced randomisation and the aligning effect due to tension in the magnetic field. Strong underlying magnetic field resists the randomization due to turbulence in interstellar conditions. The DCF method, traditionally used for the measurement of the $B$-field from dust polarization is based on the assumption that the turbulence is isotropic. However, MHD turbulence is anisotropic by nature \citep[][and references therein]{Beresnyak2019} and the DMA technique was introduced by \citet{Lazarian2022} to address this key problematic assumption of the widely used DCF method. 
	
	The DCF technique we have employed in the previous section is a modified version from the original method proposed by \citet{Houde2009}, who constrain the dispersion in the polarization angle taking into account the dispersion induced by the changes in the large scale orientation of the $B$-field as well as the beam-averaging effects. However, as discussed in \citet{Lazarian2022} the method still assumes the turbulence to be isotropic, where we use the LOS velocity dispersion as an accurate description of the POS velocity fluctuation and comparable to the observed dispersion in the polarization angle in the POS. The DMA method corrects this assumption by using the small scale structure function of the velocity centroids for the estimate of the velocity fluctuations instead of the line width velocity dispersion, similar to the method proposed by \citet{Cho2016}. The centroid velocity ($C(r)$) is defined as,
	\begin{equation}
		C(r) = \frac{\int{T_\mathrm{mb}(r,v)v dv}}{\int{T_{\mathrm{mb}}(r,v)dv}},
	\end{equation}
	where $T_{\mathrm{mb}}(r,v)$ is the brightness temperature of the CO line we have chosen with $r=(x,y)$ and $v$ its respective plane of sky position and velocity. The centroid velocities give an estimate of the average of the mean velocities from multiple individual eddies along the LOS. They are also not affected by thermal broadening, which is another parameter that contributes to the error in the $B$-field estimates from the DCF method. Using the simplest form of the DMA method from \citet{Lazarian2022} and \citet{Hu2023}, the POS $B$-field is given by;
	\begin{equation}
		B_{\mathrm{POS}}^{\mathrm{DMA}} = f \sqrt{4\pi\rho}\sqrt{\frac{D_v(\ell)}{D_{\phi}(\ell)}},
	\end{equation}
	where $D_v$ and $D_{\phi}$ are the structure functions of the velocity centroid, $C(r)$ and the polarization angle, $\phi$ within scale $\ell$, respectively. The factor $f$ is similar to the one used in the DCF method and in our case we use $f=1$ for both DCF and DMA estimates of the $B_{{}_{\mathrm{POS}}}$. The structure functions are estimated as,
	\begin{equation}
		\begin{split}
			D_v = \langle(C(r) - C(r+\ell))^2\rangle, \\
			D_{\mathrm{\phi}} = \langle(\phi(r) - \phi(r+\ell))^2\rangle.
		\end{split}
	\end{equation}
	We have applied the structure function on the same scale as used in the dispersion function for the DCF method. Thus we choose a kernel size of $w=9$, keeping the estimates consistent with the DCF from the previous section. 
	
	As discussed in \citet{Cho2016}, the DCF method gives an appropriate estimate of the $B_{{}_{\mathrm{POS}}}$ when the number of independent fluctuations (eddies) along the LOS is small. This is not the case for sightlines in the Galactic disk, as can be seen from the spectrum shown in Fig. \ref{fig:sample_spectrum}, where we clearly have multiple features along the LOS. Thus the turbulence injection and driving scale and the number of independent fluctuations along the LOS become important for the DCF method, while they are taken care of by the velocity centroid approach in the DMA method. Our approach of resolving the different individual components which contribute to the net velocity fluctuation along the LOS might address this problem with DCF to some degree. We have estimated the $B_{{}_{\mathrm{POS}}}$ for the 53 \micron\ observation from the above described DMA method using only the negative velocity part of the spectral cube (with $v<0$ km s$^{-1}$) and the corresponding map is shown in Fig. \ref{fig:B_field_s53}. The mean magnetic field estimated from the DMA method is $B_{\mathrm{POS}}^{\mathrm{DMA}}=2.79\pm2.25$ mG while that from the DCF method is $B_{\mathrm{POS}}^{\mathrm{DCF}}=2.03\pm1.8$ mG. A histogram of the distribution of $B_{{}_{\mathrm{POS}}}$ from both the methods is shown in Fig. \ref{fig:s53_DMA_hist}. Though the overall distribution of $B_{{}_{\mathrm{POS}}}$ looks similar, there are regions where there is a large difference between them. Fig. \ref{fig:s53_Bdiff_dcf_dma} shows the relation between the difference in the estimated $B_{{}_{\mathrm{POS}}}$ and the number of resolved components along the LOS ($N_{\mathrm{comps}}$) and $\sigma_v$. The difference seems to be the highest where the velocity dispersion is high and there was no apparent relation between the estimated difference and the angle dispersion. The key difference between the two methods is in the way velocity fluctuations are measured, with the velocity centroids used in DMA being a better way to constrain the turbulence driven fluctuations along the LOS. The overall agreement in the measured $B_{{}_{\mathrm{POS}}}$ is a positive indication that resolving the individual components can address some of the problems while implementing DCF for the estimate of the magnetic field. Numerical modelling of this approach paired with DMA can give further understanding on how the various velocity fluctuations along the LOS resolved from molecular spectra affect the observed polarization and will be addressed in our future work. 
	
	\section{Discussion} \label{sec:Discussion}
	\subsection{Alfv\'en Mach number ($\mathcal{M_{\mathrm{A}}}$)}
	We have used the 3D Alfv\'en Mach number ($\mathcal{M}_{\mathrm{A}}$) defined as the ratio of the turbulent velocity and the Alfv\'en speed to understand the interplay between the turbulence in the region and the magnetic field. The relation is given by,
	\begin{equation}
		\mathcal{M}_{\mathrm{A}} = \sqrt{3}\frac{\sigma_v}{\mathcal{V}_{\mathrm{A}}},
	\end{equation}
	where $\sigma_v$ is the one-dimensional non-thermal velocity dispersion which characterises the turbulence and $\mathcal{V}_{\mathrm{A}}$ is the Alfv\'en velocity given by $B/\sqrt{4\pi\rho}$, where $\rho$ is the gas mass density and $B=\sqrt{B_{{}_{\mathrm{POS}}}^{2}+B_{{}_{\mathrm{LOS}}}^{2}}$ is the total magnetic field strength which can also be approximated as $B=(4/\pi)B_{{}_{\mathrm{POS}}}$ \citep{Crutcher2004_apj}. The $\sqrt{3}$ in the equation takes care of the 3D velocity dispersion approximation from the one-dimensional estimates, assuming isotropic turbulence in the region. A value of $\mathcal{M}_{\mathrm{A}}<1$ indicates sub-Alfv\'enic condition where the magnetic field dominates the gas motion while a value of $\mathcal{M}_{\mathrm{A}}>1$ indicates the super-Alfv\'enic condition where turbulence pressure dominates over the magnetic pressure. 
	
\begin{figure}
		\centering
		\includegraphics[scale=0.4]{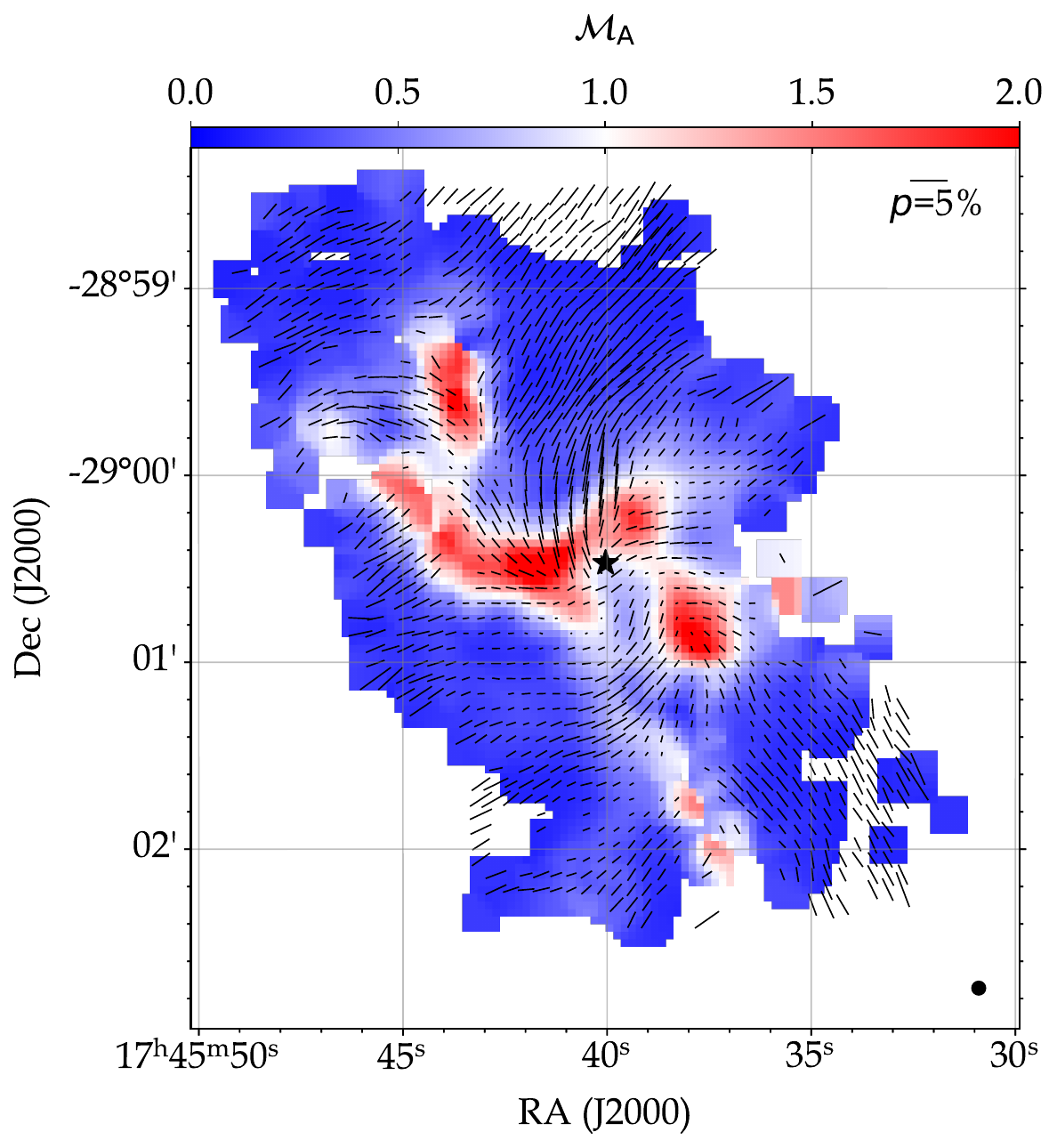}
		\caption{Alfv\'en Mach number ($\mathcal{M}_{\mathrm{A}}$) estimated for the 53 \micron\ observation. $\mathcal{M}_{\mathrm{A}}>1$ along the Eastern Arm of the minispiral indicating the gas kinematics to be driven by turbulence in the region. The HAWC+ polarization vectors are overlaid on the map with its scale shown at the top and the beam size shown at the bottom of the figure.}
		\label{fig:s53_mach}
	\end{figure}

	We have estimated $\mathcal{M}_{\mathrm{A}}$ for the HAWC+ 53 and 216 \micron\ observations. The JCMT 850 \micron\ observation was not considered due to the relatively low data quality. Though this data could be used to get a rough estimate of the field strength in the region, they are not reliable enough to draw conclusions on the importance of turbulence on these scales. The $\mathcal{M}_{\mathrm{A}}$ estimate for 216 \micron\ observation is sub-Alfv\'enic throughout the region due to its highly ordered magnetic field. The distribution of  $\mathcal{M}_{\mathrm{A}}$ is much more distinct in the 53 \micron\ observation and is shown in Fig. \ref{fig:s53_mach}. $\mathcal{M}_{\mathrm{A}}>1$ along the Eastern Arm of the minispiral where the $B_{{}_{\mathrm{POS}}}\lesssim1$mG. Turbulent pressure clearly dominates the gas motion in this region. The dust components probed by the 53 \micron\ observation show a distribution of sub- and super-Alfv\'enic conditions, in contrast to the 216 \micron\ observation where the regions is mostly sub-Alfv\'enic. \cite{Skalidis2022} studied the effect of magnetic field on the atomic-to-molecular transition in partially atomic diffuse filamentary cloud and found the turbulence to be trans-Alfv\'enic ($\mathcal{M}_{\mathrm{A}}\sim1$) in the region where the atomic-to-molecular transition takes place. The stronger magnetic field can lead to the accumulation of the atomic gas along the field lines, creating higher density sites for molecular gas formation. We observe a similar trans-Alfv\'enic region in Fig. \ref{fig:s53_mach} where $\mathcal{M}_{\mathrm{A}}\sim1$ at the white boundary which distinguishes the sub- and super-Alfv\'enic regions. The density of gas in the $\mathcal{M}_{\mathrm{A}}>1$ region is also lower than the $\mathcal{M}_{\mathrm{A}}<1$ region. $\mathcal{M}_{\mathrm{A}}>1$ is observed in the cavity of the CND, where the gas is known to be predominantly atomic \citep{Jackson1993}. Some regions in this super-Alfv\'enic part of the image also show $\mathcal{M}_{\mathrm{A}}>2$, the domain where the magnetic field can be considered to be dynamically unimportant. This could indicate the importance of the $B$-field in the formation of molecular gas in the CND. Follow-up studies with high resolution spectroscopic data, which measure the atomic and molecular composition of the CND could further our understanding of the role of the $B$-field in this transition process.
	
	\subsection{Mass-to-Flux ratio ($\lambda$)}
\begin{figure}
		\centering
		\includegraphics[scale=0.4]{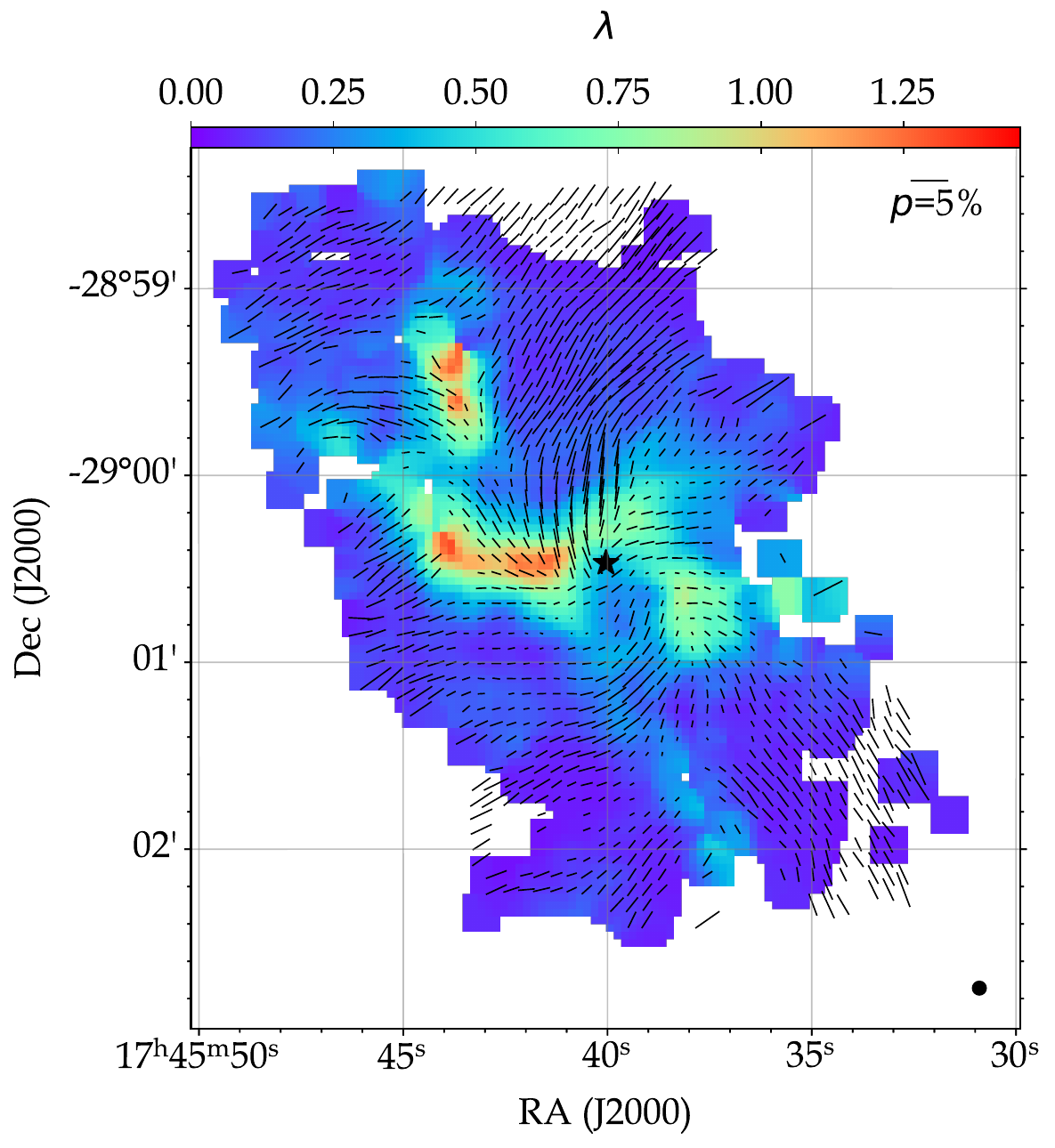}
		\caption{Mass-to-Flux ratio ($\lambda$) of the 53 \micron\ observation. $\lambda\gtrsim1$ along the Easter Arm of the minispiral within the CND where the estimated magnetic field is $B_{{}_{\mathrm{POS}}}\lesssim1$ mG. The HAWC+ polarization vectors are overlaid on the map with its scale shown at the top and the beam size shown at the bottom of the figure.}
		\label{fig:s53_lambda}
	\end{figure}

	The CND is known to be clumpy and is a key region to understand how the gas mass is being fed to our central supermassive black hole (SMBH) as well as the star formation in this region \citep{Hsieh2018,Hsieh2021}. The minispiral observed within the CND is believed to be molecular gas that are in-flowing from ambient clouds well outside the CND, and contain compact dense cores of various sizes as seen from the Atacama Large Millimeter/submillimeter Array (ALMA) observations \citep{Hsieh2019}. These cores need to be dense enough to survive the tidal disruption along the path where they become part of the CND and are eventually fed to the SMBH at the centre. An understanding of whether these cores can lead to successful star formation in this complex and highly dynamic environment around our SMBH has great implications for the study of nuclear star clusters and star formation and evolution of galaxies. One of the key parameters to probe the possible star formation in the presence of magnetic fields is the mass-to-flux ratio ($M/\phi\equiv\lambda$) which estimates whether the magnetic field can support the cloud against gravitational collapse. Following \citet{Crutcher2004_apj}, $\lambda$ is defined in terms of the critical value of mass that can be supported by magnetic flux \citep[$M_{\mathrm{crit}}=\phi_{\mathrm{crit}}/2\pi \sqrt{G}$;][]{Nakano1978} as,
	
	\begin{equation}
		\lambda = \frac{(M/\phi)_{\mathrm{obs}}}{(M/\phi)_{\mathrm{crit}}} = \frac{\mu m_{{}_{\mathrm{H}}}N(H_2)/B}{1/2\pi \sqrt{G}} = 7.6 \times 10^{-21} \frac{N(H_2)}{B},
	\end{equation}
	where $\mu=2.8$ is the mean molecular weight, $m_{\mathrm{H}}$ is the mass of hydrogen atom, $G$ is the gravitational constant, $N(H_2$) is the gas column density in cm$^{-2}$, and $B$ is the total magnetic field strength in $\mu$G. A value of $\lambda>1$ indicates that the magnetic field cannot prevent gravitational collapse of the cloud and is said to be \textit{magnetically supercritical}. If $\lambda<1$ then the cloud is magnetically supported and is said to be \textit{magnetically sub-critical}. We estimate the mass-to-flux ratio for the 53 \micron\ observation focused on the CND and the corresponding map is shown in Fig. \ref{fig:s53_lambda}. Most of the region is magnetically sub-critical due to the high magnetic field with $B_{{}_{\mathrm{POS}}}>1$ mG. Only along the Eastern Arm where we also observe $\mathcal{M}_{\mathrm{A}}>1$ do we see $\lambda\gtrsim1$ indicating the magnetic field might be not strong enough to provide support against gravity. The physical scale probed by the HAWC+ observation is not high enough to resolve the dense cores observed along the CND. However, this gives us an idea of the region of weak magnetic field, where possible star formation can be triggered in the CND.
	
	\subsection{Grain alignment and field morphology}
	From our previous study of the CND and its surroundings in \citet{Akshaya2023}, we found that the grain alignment in this region can be perfect if the dust grains contain even low levels of iron atoms (super-paramagnetic grains with $\sim20$ iron atoms per cluster). The magnetic field was assumed to be 5 mG throughout the region for the study. The results still hold with the current detailed map of the $B_{{}_{\mathrm{POS}}}$. The least magnetic field observed is within the CND with $B_{{}_{\mathrm{POS}}}\lesssim1$ mG, where we also observe a drop in the column density. Earlier Zeeman measurements of this region by \citet{Plante1995} estimate similar field strength of $B_{{}_{\mathrm{LOS}}}<1$ mG. It is interesting to note the agreement between the two methods of $B$-field estimation. Except in the few regions at the edges of the CND, the overall magnetic field for the 53 \micron\ observation shows $B_{{}_{\mathrm{POS}}}\lesssim5$ mG. Thus, we reinforce that if the dust grains in the region contain even a small percent of iron atoms, we can expect perfect alignment of dust with the magnetic field in the region where the alignment will be driven by MRAT \citep{Hoang2022AJ}. But without iron atoms, the grain alignment is only driven by radiative torques \citep{Dolginov1976,Draine1997,LazarianRAT2007} as the field in the region is not strong enough to enhance the alignment via magnetic relaxation. A detailed map of the metallicity of the region can better help constrain the degree of grain alignment and aid in the construction of a 3D  morphology of the $B$-field in this region. 
	
	\citet{Hoang2024} propose a new method to map the 3D $B$-field from the 2D dust polarization observations based on the MRAT alignment theory, taking into account the local physical conditions that affect the net polarization efficiency of the dust grains. Considering the strong magnetic field in the GC which can promote a greater degree of magnetic relaxation in contrast to the diffuse ISM, particularly in the presence of iron clusters in the dust grains \citep{Jenkins2009,Dwek2016,Altobelli2016}, it would be insightful to apply this method to derive a complete picture of the variation of the $B$-field along the LOS and how it affects the transport of material in this region.

	\subsection{3D magnetic field from multi-wavelength thermal dust polarization}
	The major focus of this work has been to map the magnetic field of the GC in multi-wavelengths and to test if regions such as the GC with multiple resolvable velocity fluctuations along the LOS can be used to create a picture of how the $B_{{}_{\mathrm{POS}}}$ changes at different depths, assuming the different wavelengths are probing different layers at varied temperatures. We believe we have achieved this to some extent given the limitations of the DCF method used to determine the magnetic field. The estimated $B_{{}_{\mathrm{POS}}}$ is in good agreement with the earlier Zeeman measurements of $B_{{}_{\mathrm{LOS}}}$ from \citet{Killeen1992} and \citet{Plante1995}. \citet{Killeen1992} obtained $B_{{}_\mathrm{LOS}}\sim2$ mG in the north and the south of the CND while \citet{Plante1995} suggested the $B_{{}_{\mathrm{LOS}}}<1$ mG within the CND. Assuming the dust polarization and Zeeman measurements trace similar regions along the CND, using our POS magnetic field from the DCF method of $B_{{}_{\mathrm{POS}}}^{\mathrm{mean}}=2.03\pm1.8$ mG, we can estimate an approximate mean full strength of 3D $B$-field of the CND as;
	\begin{eqnarray}
		B=\sqrt{B_{{}_{\mathrm{POS}}}^{2}+B_{{}_{\mathrm{LOS}}}^{2}},
	\end{eqnarray}
	which yields $B\sim2.85$ mG. The observed polarization in the 53 \micron\ appears to be independent of the deeper velocity structures identified as shown in Fig. \ref{fig:acorns}, but the 216 and 850 \micron\ observations seem to trace the relatively cooler dust from different depths along the LOS. 
	
	\subsection{Implications of B-field strength from multi-wavelength polarization at the GC}
\begin{figure}
		\centering
		\includegraphics[scale=0.45]{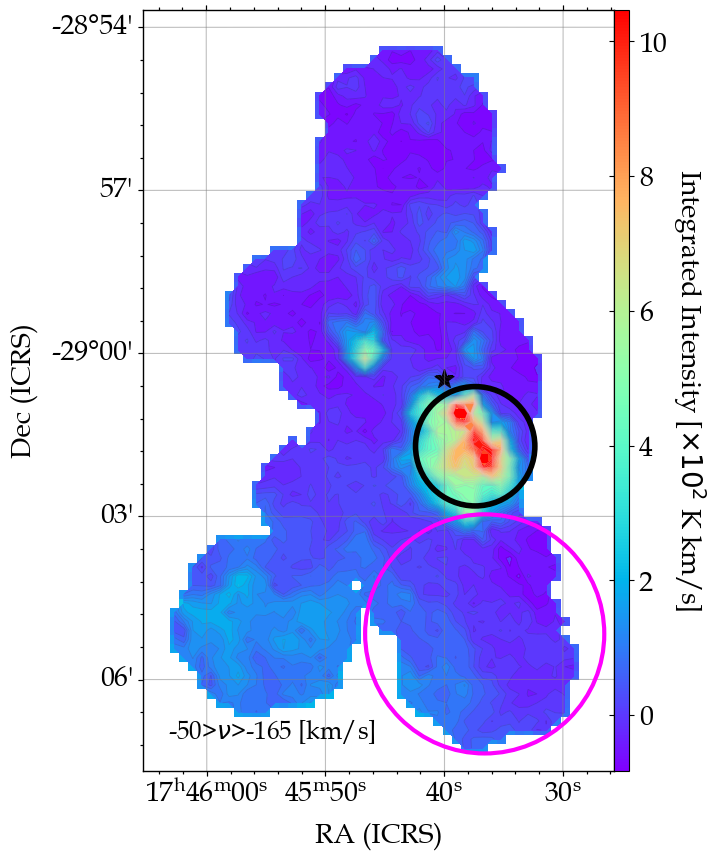}
		\caption{The NLE (black circle) observed in CO ($J=3\rightarrow2$) spectra evident in the region covered by the JCMT/SCUPOL observation, right below the CND and above the location of the 20 km s$^{-1}$ cloud (shown as the magenta circle). The component appears in the negative velocity range of $-50>v>-165$ km s$^{-1}$.}
		\label{fig:CO_nle}
	\end{figure}

	The CND is well known to be in the influence of the gravitational potential of the Sgr A$^*$ and also play a role in the accretion of material onto the inner sub-parsec scale ionized cavity surrounding Sgr A$^*$ \citep{Solanki2023}. There is also evidence of collisional interaction between the 20 km s$^{-1}$ cloud and the outer edges of the CND, which might aid further mass accretion onto the CND and the inner cavity \citep{Takekawa2017}. The evolution studies of the CND suggest its formation due to the interaction and break down of giant molecular clouds with the gravitational potential of the Sgr A$^*$, with two molecular clouds (20 km s$^{-1}$ and 50 km s$^{-1}$ clouds) ideally located at the Galactic south of the CND \citep{Sanders1998,Oka2011,Mapelli2016}. The CND is also observed to have an extended feature in the negative Galactic longitude direction called the negative-longitude extension \citep[NLE;][]{Serabyn1986,Sutton1990,Takekawa2017}. This is a foreground (with respect to the CND) feature and we observe a similar structure in the velocity range of $-50>v>-165$ km s$^{-1}$ shown in Fig. \ref{fig:CO_nle}. Considering the similar $B_{{}_{\mathrm{POS}}}$ estimates for the 53 \micron\ observation covering the CND and the 850 \micron\ observation with the CND, NLE, and the 20  km s$^{-1}$ features, the 850 \micron\ maybe tracing the CND and its foreground material at a lower temperature. This cannot be confirmed due to the low data quality of the SCUPOL observation. However, it is interesting nonetheless to disentangle the location of the magnetic field measured at 850 \micron\ to get a view of how it varies with the depth along the LOS. The 216 \micron\ observations measure the highest magnetic field out of the three observations that we have considered. From Fig. \ref{fig:acorns} it is evident that there is some degree of interaction between the cloud structures in the negative velocity region but there is no interaction between the $v<0$ km s$^{-1}$ and $v>0$ km  s$^{-1}$ cloud features. Also by looking at Fig. \ref{fig:s216_components}, the majority of the emission in the 216 \micron\ observation matches well with the component morphologies at $v>0$ km s$^{-1}$, unlike the 53 \micron\ intensity which matches well with the integrated morphologies from $v<0$ km s$^{-1}$. If the observed dust emissions are arising from these proposed components, then the strong magnetic field of the 216 \micron\ observation can be attributed to the velocity components in the background with respect to the CND. The uniformity in the polarization vectors at this wavelength has been noted by earlier observation at 240 \micron\ by the PILOT experiment as well \citep{Mangilli2019}. Considering our data quality check and the agreement between the observations taken by other instruments and in different modes, this uniformity of polarization can be treated as a real physical effect, arising due to dust grains being perfectly aligned with a strong magnetic field. 
	
	Modelling and numerical simulations are necessary to further confirm this idea that the different field strengths observed from multi-wavelength polarization could be a result of dust being traced at different depths along the LOS and will be addressed in our future work. The growth of the CND and subsequent feeding of material into the central black hole is an ongoing study with most of the simulation not yet considering the effect of the magnetic field in the material dynamics due to the complexity of the region. Recent observations reveal the clumpy nature of the CND with densities ranging from $10^3-10^8$ cm$^{-3}$ \citep{Vollmer2001,Vollmer2002,Requena-Torres2012,Hsieh2021}. Our observations are only able to probe the density range of $\sim10^4$ cm$^{-3}$. The turbulence of the high density clumps might lead to a tangled magnetic field that is lost within the beam size of our observations. As a consequence the measured $B$-field might be underestimated and trace the low density gas of the CND. Observations at higher spatial resolution are necessary to resolve the effect of the clumpy nature of the CND on its observed magnetic field. An overview of the strength and 3D morphology of the $B$-field in the region that can be estimated from polarization observation, complemented with the current grain alignment physics is a great way to further our understanding of the material transport at the centre of our Galaxy.
	
	\section{Summary and Outlook} \label{sec:Summary}
	We have used thermal dust polarization observations at 53, 216, and 850 \micron\ combined with the spectrum of the CO ($J=3\rightarrow2$) transition to map the POS magnetic field for a region of about 30 pc around the centre of our Galaxy. The main results of our study are as follows;
	\begin{enumerate}[label=\arabic*.,leftmargin=*]
		\item The velocity dispersion in the region can be decomposed into multiple physically distinct components with an average dispersion of about $\sim9$ km s$^{-1}$.
		\item The physical morphologies of the negative velocity structures match best with the observed morphology of the CND and the minispiral at 53 \micron, indicating that these structures are at a much higher temperature than those at positive velocities. This also shows that only the negative velocity components contribute to the velocity dispersion of the gas associated with the dust emission observed at this wavelength.
		\item We used the DCF method to estimate the map of the $B_{{}_{\mathrm{POS}}}$ for all the three observations and found the mean field to be $2.03\pm1.8$, $6.76\pm3.49$, and $1.43\pm0.57$ mG at 53, 216, and 850 \micron\ respectively. 
		\item Most of the region encompassing the CND and the minispiral is sub-Alfv\'enic with $\mathcal{M}_{\mathrm{A}}<1$ except along the Eastern Arm of the minispiral where $\mathcal{M}_{\mathrm{A}}>1$, indicating the gas motion being driven by turbulence.
		\item The Eastern Arm also has a mass-to-flux ratio of $\lambda\gtrsim1$ indicating magnetic field cannot prevent gravitational collapse. The CND is known to have several clumps with $n_{{}_{\mathrm{H}}}>10^7$ cm$^{-3}$. Although these scales are not probed in the current observation, the map of $\lambda$ shows the likely region for star formation in the CND.
		\item We find good agreement between the $B_{{}_{\mathrm{POS}}}$ estimated from DCF and DMA method for the 53 \micron\ observation, indicating that the decomposition of the spectra into their velocity components might overcome the drawback of the velocity dispersion used in the DCF method, where determining the turbulence driving scale and the number of independent fluctuations along the LOS become crucial to overcome the overestimation of the $B$-field.
		\item The similarity in the estimated $B_{{}_{\mathrm{POS}}}$ for the 53 \micron\ and the 850 \micron\ observation might be due to dust emission in these wavebands tracing the same components, mostly including the CND and its foreground, whereas the 216 \micron\ emission might mainly originate from a component behind the CND, with a much stronger magnetic field.
	\end{enumerate}
	
	Further studies with numerical simulations can investigate if combining such multi-wavelength polarization observations with the information of the 3D distribution of dust along the LOS can be used to create a 3D morphology of the magnetic field. As an extension of this work, we plan to use the recently upgraded POLArized RadIation Simulator \citep[POLARIS;][]{Reissl2016} by \citet{GiangPolaris2022}, which incorporates the latest grain alignment theories discussed in our previous work \citep{Akshaya2023} to model how the distribution of independent velocity fluctuations affect the observed polarized emission at different wavelengths in this region. Combined with the limited $B_{{}_{\mathrm{LOS}}}$ Zeeman measurements, this might be a step closer to getting a comprehensive view of the elusive 3D morphology of the magnetic field in the Galactic disk. Due to the nature of the grain alignment and how sensitive it is to the local physical conditions like the metallicity, temperature, dust composition, local density, and the amount of incident radiation, we need to use the latest knowledge of the dust alignment physics to get a full picture of the dynamical interaction between the magnetic field and the material in this complex region.
	
	
	\section*{Acknowledgements}
	We thank the anonymous reviewer for their useful comments that helped improve this paper.  MSA thanks Dr. Lopez-Rodriguez and Dr. G. S. Pillai for the insightful discussion on the polarization data quality assessment. This work was partly supported by a grant from the Simons Foundation to IFIRSE, ICISE (916424, N.H.). This study is based in part on observations made with the NASA/DLR Stratospheric Observatory for Infrared Astronomy (SOFIA). SOFIA is jointly operated by the Universities Space Research Association, Inc. (USRA), under NASA contract NNA17BF53C, and the Deutsches SOFIA Institut (DSI) under DLR contract 50 OK 2002 to the University of Stuttgart. This work made use of Astropy:\footnote{http://www.astropy.org} a community-developed core Python package and an ecosystem of tools and resources for astronomy \citep{astropy:2013, astropy:2018, astropy:2022}. This research made use of APLpy, an open-source plotting package for Python \citep{aplpy}.
	
	\section*{Data Availability}
	The data underlying this article will be shared on reasonable request to the corresponding author.
	
	
	\bibliographystyle{mnras}
	\bibliography{references} 
	
	
	\appendix
	\section{Supplementary Figures}
	\begin{figure*}
		\includegraphics[scale=0.45]{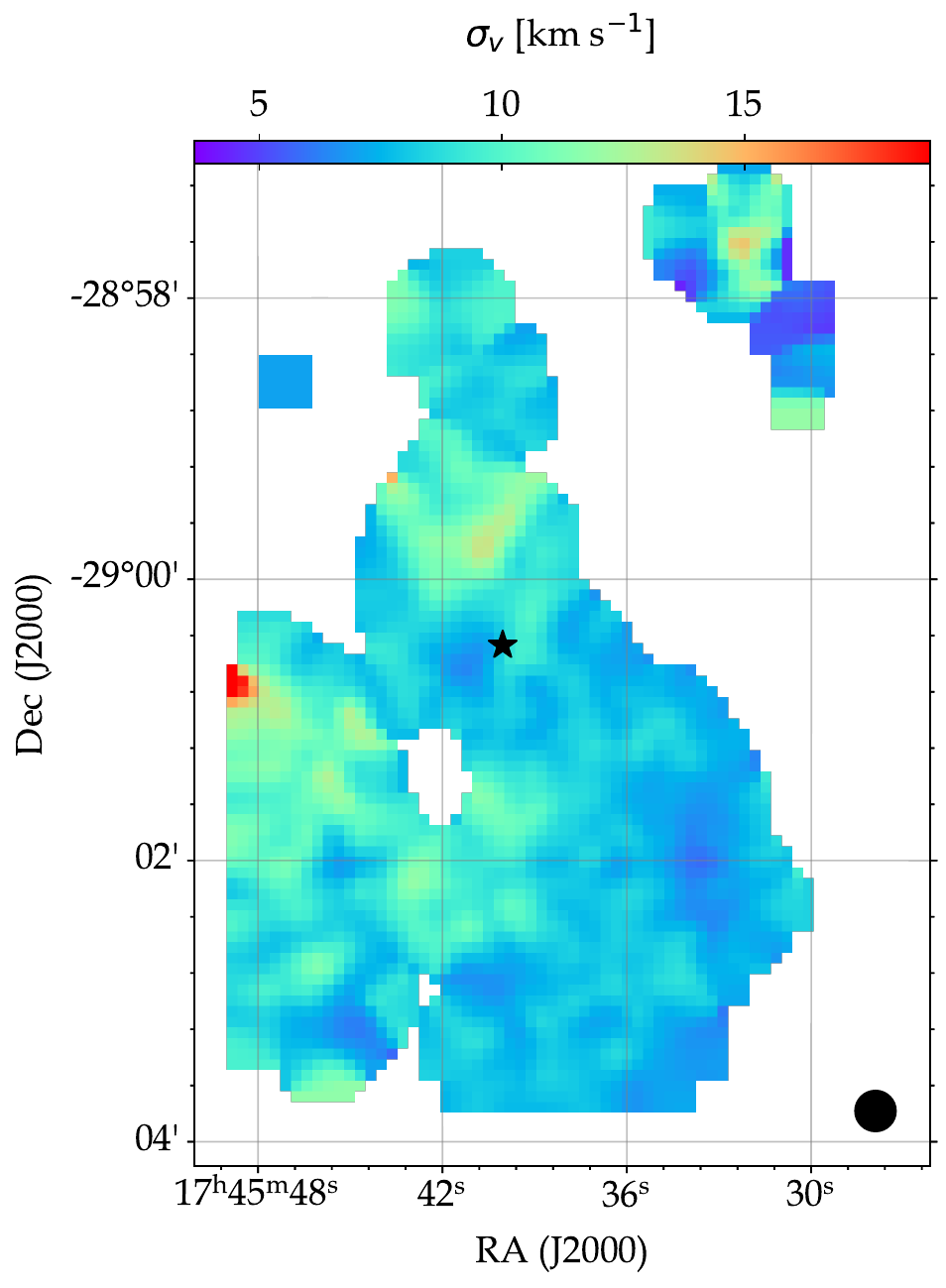}
		\hspace{1cm}
		\includegraphics[scale=0.45]{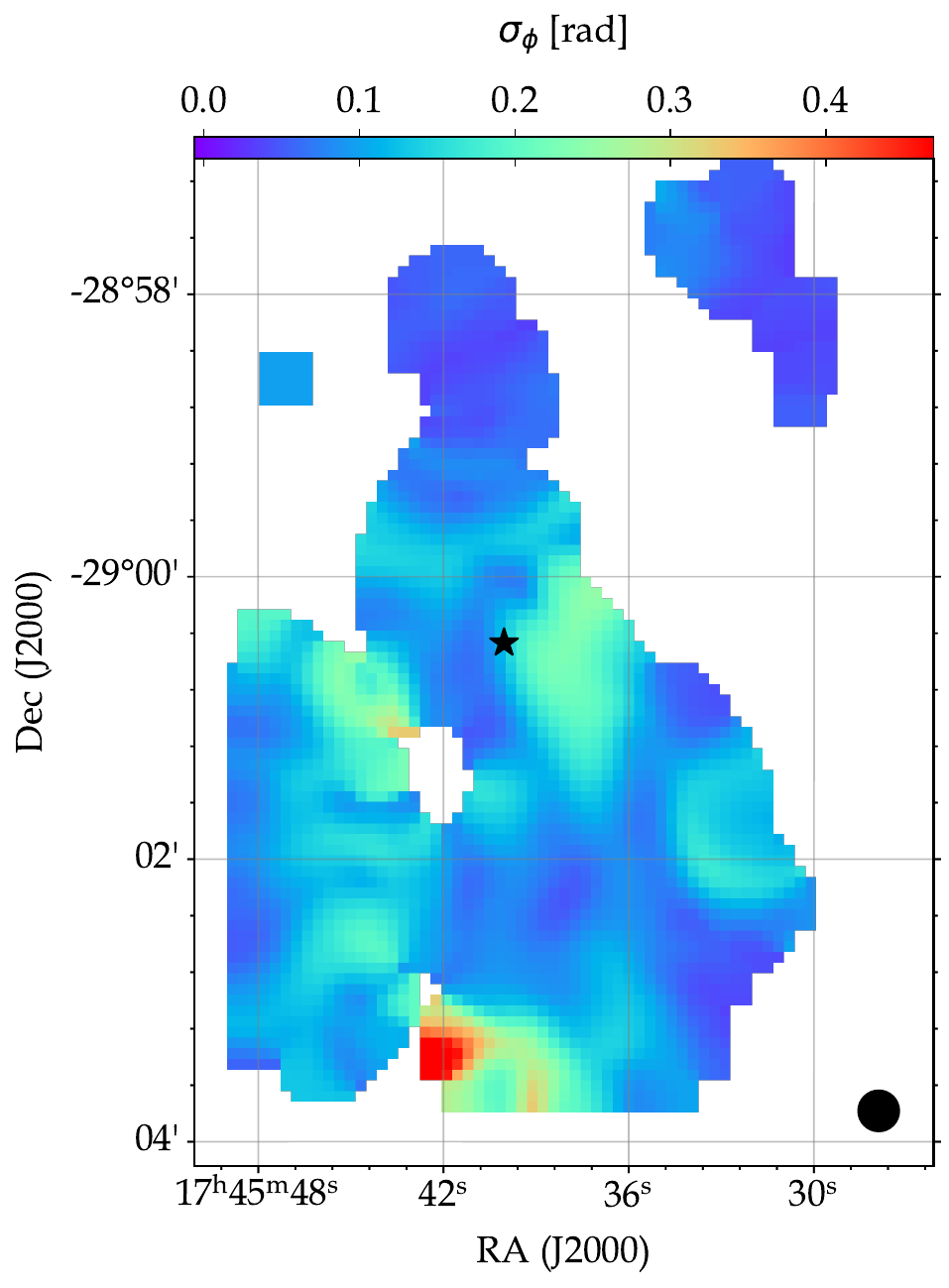}
		\caption{Maps of the dispersion in velocity (left) and polarization angle (right) used for the estimation of the $B_{{}_{\mathrm{POS}}}$ from the DCF method for the HAWC+ 216 \micron\ observation.}
		\label{fig:s216_V_phi}
	\end{figure*}
	
	\begin{figure*}
		\includegraphics[scale=0.45]{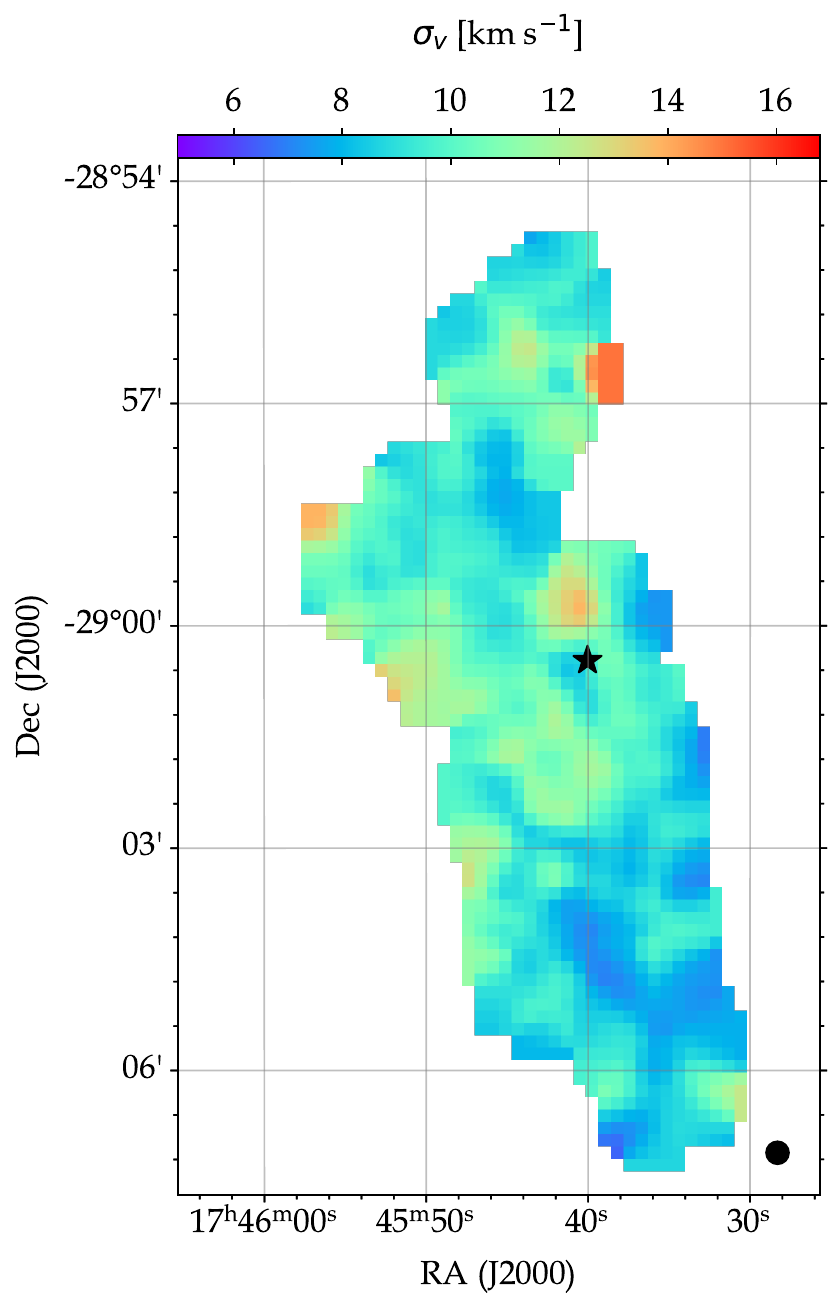}
		\hspace{1cm}
		\includegraphics[scale=0.45]{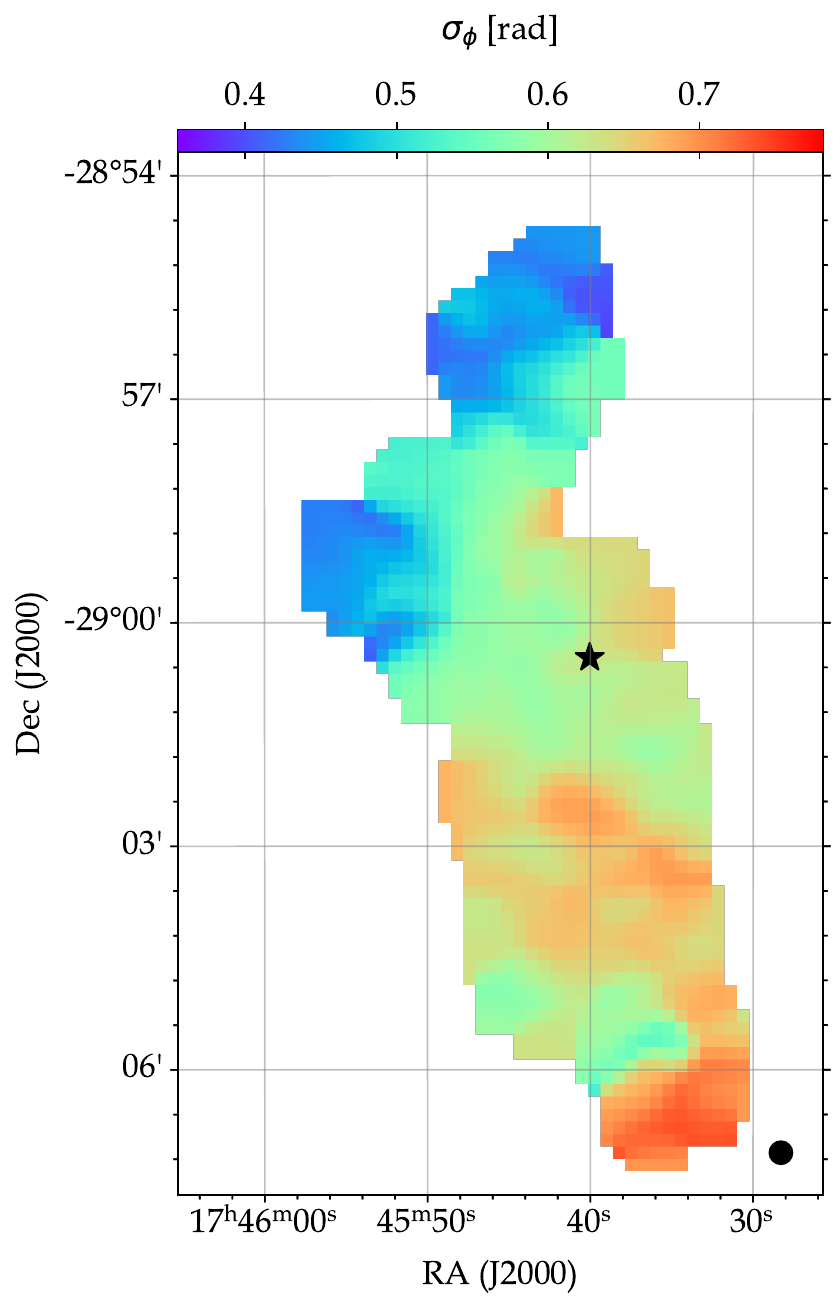}
		\caption{Maps of the dispersion in velocity (left) and polarization angle (right) used for the estimation of the $B_{{}_{\mathrm{POS}}}$ from the DCF method for the SCUPOL 850 \micron\ observation.}
		\label{fig:j850_V_phi}
	\end{figure*}

	\bsp	
	\label{lastpage}
\end{document}